\documentclass[AMA, STIX1COL]{WileyNJD-v2}

\articletype{Research Article}%

\received{11 August, 2022}
\revised{6 April, 2023}
\accepted{20 April, 2023}
\doi{https://doi.org/10.1002/spe.3214}

\usepackage{amsmath,amsfonts}
\usepackage{float}
\usepackage{graphicx}
\usepackage{hyperref}
\usepackage{textcomp}
\usepackage{pgf}
\usepackage{xcolor}
\usepackage{multirow}
\usepackage{xspace}
\usepackage{tikz}
\usepackage{comment}
\usetikzlibrary{decorations.pathreplacing}
\usepackage{paralist}
\usepackage[shortlabels]{enumitem}
\usepackage[utf8]{inputenc}
\usepackage{listings}
\usepackage{caption}
\usepackage{lst-langs-styles/llvm/lang}
\usepackage{lst-langs-styles/nasm/style}
\usepackage{algorithm}
\usepackage{amsthm}
\usepackage{subcaption}
\usepackage{array}
\usepackage{ulem}
\usepackage{makecell}
\usepackage{microtype}

\def\BibTeX{{\rm B\kern-.05em{\sc i\kern-.025em b}\kern-.08em
    T\kern-.1667em\lower.7ex\hbox{E}\kern-.125emX}}

\lstdefinestyle{customc}{
  language=C,
  basicstyle=\ttfamily\footnotesize,
  keywordstyle=\sffamily\bfseries,
  keywordstyle=[2]\color{red}\bfseries,
  stringstyle=\ttfamily,
  numbers=left,
  numbersep=5pt,
  xleftmargin=.12\columnwidth,
  morekeywords={match, m_FSub, m_FNeg, m_Value},
  tabsize=2,
  breaklines=true,
  morecomment=[l]{\#},
}

\algrenewcommand\algorithmicindent{0.8em}
\algnewcommand{\LineComment}[1]{\State \(\triangleright\) #1}

\newcolumntype{M}[1]{>{\centering\arraybackslash}m{#1}}

\def\REG{\textsuperscript{\textregistered}}
\def\TM{\textsuperscript{\texttrademark}}
\def\Intel{Intel\xspace}
\def\IntelREG{Intel\REG\xspace}
\def\AMD{AMD\xspace}
\def\AMDREG{AMD\REG\xspace}
\def\IBMREG{IBM\REG\xspace}
\def\IBM{IBM\xspace}
\def\XeonTM{Xeon\TM\xspace}

\def\EpycTM{EPYC\TM\xspace}

\def\PowerNineTM{POWER9\TM\xspace}
\def\PowerNine{POWER9\xspace}
\def\PowerTenTM{POWER10\TM\xspace}
\def\PowerTen{POWER10\xspace}
\def\SVE{Scalable Vector Extension\xspace}
\def\ArmREG{Arm\REG\xspace}
\def\Arm{Arm\xspace}
\def\CascadeLake{Cascade Lake\TM\xspace}
\def\ZenTwo{Zen 2\TM\xspace}
\def\PowerPC{PowerPC\TM\xspace}

\newcommand{\rsec}[1]{Section~\ref{sec:#1}}
\newcommand{\rtab}[1]{Table~\ref{tab:#1}}
\newcommand{\rfig}[1]{Figure~\ref{fig:#1}}
\newcommand{\rlst}[1]{Listing~\ref{lst:#1}}
\newcommand{\ralg}[1]{Algorithm~\ref{alg:#1}}

\makeatletter
\let\OldStatex\Statex
\renewcommand{\Statex}[1][3]{%
  \setlength\@tempdima{\algorithmicindent}%
  \OldStatex\hskip\dimexpr#1\@tempdima\relax}
\makeatother

\newif\ifcomments
\commentstrue 

\newcommand{\bk}[1]{{\ifcomments\small\sf \color{brown} ([Braedy]: #1)\fi}}
\newcommand{\jc}[1]{{\ifcomments\small\sf \color{orange} ([João]: #1)\fi}}

\newcommand{\rem}[1]{}

\newcommand{\gemm}{{GEMM}\xspace}
\newcommand{\gemms}{{GEMMs}\xspace}
\newcommand{\syrtwok}{{SYR2K}\xspace}
\newcommand{\sgemm}{{SGEMM}\xspace}
\newcommand{\sgemms}{{SGEMMs}\xspace}

\newcommand{\code}[1]{\lstinline[basicstyle=\ttfamily,breaklines=true]|#1|}

\newcommand{\mc}{\texttt{mc}\xspace}
\newcommand{\kc}{\texttt{kc}\xspace}
\newcommand{\nc}{\texttt{nc}\xspace}
\newcommand{\mr}{\texttt{mr}\xspace}
\newcommand{\kr}{\texttt{kr}\xspace}
\newcommand{\nr}{\texttt{nr}\xspace}
\newcommand{\VL}{\texttt{VL}\xspace}
\newcommand{\kl}{\texttt{kl}\xspace}

\newcommand{\Intrinsic}{\texttt{Intrinsic}\xspace}
\newcommand{\Tiling}{\texttt{Tiling}\xspace}
\newcommand{\TilingPacking}{\texttt{Tiling+Packing}\xspace}
\newcommand{\BLAS}{\texttt{BLAS}\xspace}
\newcommand{\EIGEN}{\texttt{Eigen}\xspace}
\newcommand{\PLUTO}{\texttt{PLuTo}\xspace}

\newcommand{\TLB}{{Translation Lookaside Buffer}\xspace}

\begin{document}

\title{Fast Matrix Multiplication via Compiler-only Layered Data Reorganization and Intrinsic Lowering}

\author[1]{Braedy Kuzma}

\author[1]{Ivan Korostelev}

\author[1]{João P. L. de Carvalho*}

\author[2]{José E. Moreira}

\author[3]{Christopher Barton}

\author[4]{Guido Araujo}

\author[1]{José Nelson Amaral}

\authormark{Braedy Kuzma \textsc{et al}}

\address[1]{%
  \orgdiv{Computing Science Department},%
  \orgname{University of Alberta},%
  \orgaddress{\state{Alberta},%
  \country{Canada}}%
}

\address[2]{%
  \orgdiv{Thomas J. Watson Research Center},%
  \orgname{IBM Corporation},%
  \orgaddress{\state{New York},%
  \country{USA}}%
}

\address[3]{%
  \orgdiv{IBM Canada Software Laboratory},%
  \orgname{IBM Corporation},%
  \orgaddress{\state{Ontario},%
  \country{Canada}}%
}

\address[4]{%
  \orgdiv{Institute of Computing},%
  \orgname{UNICAMP},%
  \orgaddress{\state{São Paulo},%
  \country{Brazil}}%
}

\corres{*João P. L. de Carvalho, 8900 114 St NW, Edmonton, AB. \email{joao.carvalho@ualberta.com}}

\presentaddress{Computing Science Department, University of Alberta, 8900 114 St NW, Edmonton, AB}

\jnlcitation{\cname{%
\author{Braedy Kuzma},
\author{Ivan Korostelev},
\author{João P. L. de Carvalho},
\author{José Moreira},
\author{Christopher Barton}, and
\author{José Nelson Amaral}} (\cyear{2022}),
\ctitle{Fast Matrix Multiplication via Compiler-only Layered Data Reorganization and Intrinsic Lowering}, \cjournal{Journal of Software: Practice and Experience}, \cvol{2022;XX:X--X}.}

\abstract[Summary]{
The resurgence of machine learning has increased the demand for high-performance basic linear algebra subroutines (BLAS), which have long depended on libraries to achieve peak performance on commodity hardware.
High-performance BLAS implementations rely on a layered approach that consists of tiling and packing layers --- for data (re)organization --- and micro kernels that perform the actual computations.
The algorithm for the tiling and packing layers is target independent but is parameterized to the memory hierarchy and register-file size.
The creation of high-performance micro kernels requires significant development effort to write tailored assembly code for each architecture.
This hand optimization task is complicated by the recent introduction of matrix engines by \IBMREG's \PowerTenTM (Matrix Multiply Assist -- MMA), \IntelREG (Advanced Matrix eXtensions -- AMX), and \ArmREG (Matrix Extensions -- ME) to deliver high-performance matrix operations.
This paper presents a compiler-only alternative to the use of high-performance libraries by incorporating, to the best of our knowledge and for the first time, the automatic generation of the layered approach into LLVM, a production compiler.
Modular design of the algorithm, such as the use of  LLVM's matrix-multiply intrinsic for a clear interface between the tiling and packing layers and the micro kernel, makes it easy to retarget the code generation to multiple accelerators.
The parameterization of the tiling and packing layers is demonstrated in the generation of code for the MMA unit on IBM's \PowerTen.
This paper also describes an algorithm that lowers the matrix-multiply intrinsic to the MMA unit.
The use of intrinsics enables a comprehensive performance study.
In processors without hardware matrix engines, the tiling and packing delivers performance up to $22\times$ (\Intel) --- for small matrices --- and more than $6\times$ (\PowerNine) --- for large matrices --- faster than PLuTo, a widely used polyhedral optimizer.
The performance also approaches high-performance libraries and is only $34\%$ slower than OpenBLAS and on-par with Eigen for large matrices.
With MMA in \PowerTen this solution is, for large matrices, over $2.6\times$ faster the vector-extension solution, matches Eigen performance, and achieves up to $96\%$ of \BLAS peak performance.
}


\maketitle

\section{Introduction}

New machine-learning algorithms, combined with the ever increasing demands of scientific and business analytics applications, highlight the importance of improving the performance of matrix  algorithms and matrix multiplication in particular~\cite{2017vasudevan}.
General Matrix Multiplication (\gemm) is a routine heavily used in high-performance computing and neural networks~\cite{waugh2020use},  both as a standalone operation and as a crucial component of other linear-algebra algorithms, such as LU decomposition.
As the typical matrix sizes in \gemm operations for deep-learning workloads approach the order of 10,000~\cite{sigma2020}, efficiently partitioning the matrices to fit into the cache hierarchy is key to performance.
The state-of-the-art approach in high-performance numerical libraries uses a layered approach for matrix multiplication that consists of (re)organizing the data to improve data locality as it moves from the main memory through the memory hierarchy and then relying on specialized assembly code to execute the multiplications efficiently in each targeted architecture.

This layered method, explained by Goto and Geijn~\cite{goto2008} and used in both  proprietary (e.g. \Intel MKL, \IBM ESSL)  and open-source (e.g. OpenBLAS~\cite{2012xianyi}, BLIS~\cite{blis2015}, Eigen~\cite{eigenweb}) highly-optimized libraries, consists of a two-layer approach.
First, a \textit{macro kernel} performs tiling and packing of the operand matrices  across the caches.
In a second layer, a \textit{micro kernel}, implemented by means of compiler \textit{builtins} or direct assembly instructions, extracts blocks from the packed tiles and executes the block multiplication.
This work contributes new ideas to both the macro and micro kernels.

Libraries have been successful in exploiting the memory hierarchy to perform efficient matrix operations.
However, they all share some  drawbacks:
\begin{inparaenum}
	\item users must download and install architecture-specific libraries;
	\item each library needs to be tailored by writing assembly code for every new architecture design;
	\item manual changes to the users' code are necessary to call the libraries; and
	\item often there is a time lag between the introduction of a new architecture and the creation of a specialized micro kernel for that architecture in a library.
\end{inparaenum}
Overcoming these drawbacks would lead to broader utilization of the layered approach and of specialized micro kernels~\cite{waugh2020use}.

Linear-algebra libraries share similar code for matrix multiplication because they all make use of the ideas described by Goto and Geijn~\cite{goto2008}.
This insight leads to the first contribution of this paper, which aims at capturing the layered strategy described by Goto and Geijn in a compiler-only LLVM-based optimization pass.
Implementing the layered strategy  as a general-purpose compiler pass brings three benefits.
\begin{inparaenum}
	\item programs written in all languages that are supported by the LLVM frontend (e.g., C, Fortran, Go, Rust) can leverage the strategy when there is no library interface or a library implementation is outdated;
	\item the tiling and packing code, which is common to all libraries, is automatically generated by the new compiler pass;
	\item the algorithm to lower the LLVM intrinsic to architecture-specific micro-kernel code only needs to be implemented once for each architecture.
\end{inparaenum}

The resurgence of machine learning also led to the introduction of application-specific accelerators into general-purpose CPUs.
Examples include \Intel's Advanced Matrix Extension (AMX)~\cite{IntelISA}, \Arm's Matrix Extension (ME)~\cite{ArmISA} and IBM \PowerTen's Matrix Multiply Assist (MMA)~\cite{9352481,PowerISA}.
AMX is an off-core accelerator with a dedicated register file that employs inner product operations to compute in-register matrix multiply using novel tile registers~\cite{IntelISA}.
New instructions allow the CPU to communicate with AMX through an accelerator command queue~\cite{IntelISA}.
Both Arm's ME and IBM's MMA unit are on-core extensions that use SIMD vector registers for input and output operands.
However, ME and AMX rely on inner products for multiplication, while MMA uses outer products.

This heterogeneity of accelerators further complicates the hand optimization of micro kernels that library developers must perform.
In a compiler-only code generation path, the adoption of LLVM's intermediate representation (IR) \code{llvm.matrix.multiply} intrinsic abstracts target-specific operations under a clear interface reducing the need for specialized micro kernels to a single implementation of the intrinsic.
This LLVM intrinsic computes the product of two fixed-size matrices.
LLVM provides a generic lowering algorithm that unrolls the matrix-multiply computations to target-independent IR code.
LLVM's backends can then further lower IR code to target-specific machine code for any of the many backends supported.
For instance, another contribution of this paper is a lowering algorithm of the LLVM \code{llvm.matrix.multiply} intrinsic that efficiently utilizes the new MMA unit in \PowerTen.
When generating code for MMA this intrinsic computes a fast outer-product-based matrix multiplication for the micro kernel.

Summarizing, this paper makes the following contributions:
\begin{itemize}
  \item
    An algorithm for a compiler-only, architecture-independent, tiling \& packing strategy for the macro kernel that improves upon the strategy described by Goto et al.~\cite{goto2008} (\rsec{macro-level-algorithm}).
    The incorporation of this algorithm in LLVM leverages all available backends for processor architectures by building upon a compiler-intrinsic micro kernel instead of a hand-crafted assembly micro kernel as in most high-performance BLAS libraries;
  \item
    An algorithm to lower the LLVM \code{llvm.matrix.multiply} intrinsic (micro kernel) to the new IBM MMA extension (\rsec{microLevel}).
    The specialized code generated for MMA benefits not only the code generated by the macro-level algorithm, but any compilation path that uses the intrinsic.
  \item
    A thorough experimental evaluation that shows that:
    \begin{inparaenum}
      \item the proposed macro-level algorithm, even when coupled with a generic intrinsic lowering, can perform more than $22\times$ faster -- for small matrices on \Intel -- and more than $6\times$ faster -- for large matrices on \PowerNineTM -- than PLuTO\cite{pluto}, a widely used polyhedral optimizer.
      \item this compiler-only approach generates code that is on-par with \EIGEN and $34\%$ slower than BLAS on \PowerNine;
      \item coupling the macro-level algorithm with an MMA-specific lowering of the \code{llvm.matrix.multiply} intrinsic in \PowerTen achieves:
         \begin{inparaenum}
             \item more than $2.6\times$ the performance of the VSX micro kernel;
             \item $10\%$ more performance than \BLAS for a small \sgemm and up to $96\%$ of \BLAS peak performance for large \sgemm;
             \item $83\%$ faster code than \EIGEN for large matrices.
         \end{inparaenum}
    \end{inparaenum}
\end{itemize}

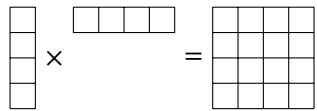
\begin{figure}[t]

    \centering
    \begin{tikzpicture}[scale=1/3]
      \draw[step=1, shift={(0, 0)}] (0, 0) grid +(1, 4);
      \node at (1.75, 2) {$\times$};
      \draw[step=1, shift={(2.5, 3)}] (0, 0) grid +(4, 1);
      \node at (7.25, 2) {$=$};
      \draw[step=1, shift={(8, 0)}] (0, 0) grid +(4, 4);
    \end{tikzpicture}
	\caption{Outer-product (rank-$1$ update) operation.}
    \label{fig:outerProduct}
  \vspace{-0.15cm}
\end{figure}

An overview of the \PowerTen Matrix-Multiply Assist (MMA) facility follows in \rsec{MMA}.
Then, \rsec{CodeGen} describes the proposed code generation approach, detailing the algorithms for the macro and micro kernels.
\rsec{results} presents and analyzes the experimental results.
\rsec{relatedWork} describes and puts in perspective the works related to this paper, and finally, \rsec{conclusion} presents conclusions.

\section{Matrix-Multiply Assist in \PowerTen}
\label{sec:MMA}

The Matrix-Multiply Assist (MMA) instructions were introduced in PowerISA 3.1~\cite{PpcISA} as an extension of the Vector-Scalar Extension (VSX) facility.
IBM's \PowerTen processor is the first to implement these new instructions and functional units that perform two-dimensional matrix operations.
MMA relies on 512-bit accumulator (ACC) registers to represent matrices, which can be manipulated by BLAS-like rank-$k$ operations that consume vector registers as inputs.
Each accumulator register is associated with four of the architecture's 128-bit vector-scalar registers (VSRs).
While an accumulator is being used for MMA instructions, the associated VSRs are blocked from use.
Up to eight of these accumulators can be use simultaneously, leaving 32 of 64 VSRs available for use as vector registers.


Matrix multiplication and other linear algebra operations can be expressed as a series of rank-$k$ update operations.
These operations compute the product of an $m \times k$ matrix by another $k \times n$ matrix, accumulating the result into an $m \times n$ matrix.
When $k = 1$, the operation reduces to an outer product of two vectors, of $m$ and $n$ elements, respectively as illustrated in \rfig{outerProduct}.
A rank-$k$ update can itself be decomposed to a sequence of $k$ outer products.

Outer-products have high-computational density since they are two-dimensional operations that compute $mn$ element-wise operations from $m + n$ input values.
Therefore, they are the standard building block of high-performance linear algebra frameworks such as OpenBLAS and Eigen.
In processors with one-dimensional vector instructions, the outer products are emulated using a combination of splatting and element-wise multiply-add instructions.
MMA bypasses this emulation step by directly supporting outer product.

In the MMA rank-$k$ update instructions, the updated matrix is stored in an ACC, while the operand vectors (or matrices) are provided through VSRs.
The result matrix is either a $4 \times 4$ matrix of 32-bit elements (floating-point or integer) or a $4 \times 2$ matrix of 64-bit floating point elements.
The $k$ is a function of the input data type, which can vary from 4-bit integers ($k = 8$) to 32- or 64-bit floating-point numbers ($k = 1$).
For all input data types of 32-bit or less, the multiplying operands are represented by one VSR each.
For 64-bit inputs, one operand is represented by a pair of VSRs ($4 \times 64$-bit elements) and the other by a single VSR ($2 \times 64$-bit elements).
A summary of the MMA rank-$k$ update instructions is shown in Table~\ref{tab:mmaInsts}.

\begin{table}[t]
  \centering
  \caption{MMA instruction summary.}
  \vspace{-0.25cm}
  \begin{tabular}{| c | c | c |}
    \hline
    Input type 				& Computation size 		& Result 							\\
	  				& $m \times k \cdot k \times n$	& shape and type						\\ 	\hline
    4-bit integer (\code{i4}) 		& $4 \times 8 \cdot 8 \times 4$ & \multirow{3}{*}{\shortstack[c]{$4 \times 4$\\\code{i32}}} 	\\ 	\cline{1-2}
    8-bit integer (\code{i8}) 		& $4 \times 4 \cdot 4 \times 4$ & 								\\ 	\cline{1-2}
    16-bit integer (\code{i16}) 	& $4 \times 2 \cdot 2 \times 4$ & 								\\ 	\hline
    brain-float (\code{bf16}) 		& $4 \times 2 \cdot 2 \times 4$ & \multirow{3}{*}{\shortstack[c]{$4 \times 4$\\\code{f32}}} 	\\ 	\cline{1-2}
    IEEE half-precision (\code{f16}) 	& $4 \times 2 \cdot 2 \times 4$ & 								\\ 	\cline{1-2}
    IEEE single-precision (\code{f32}) 	& $4 \times 1 \cdot 1 \times 4$ & 								\\ 	\hline
    IEEE double-precision (\code{f64}) 	& $4 \times 1 \cdot 1 \times 2$ & \begin{tabular}{c}$4 \times 2$\\\code{f64}\end{tabular} 	\\ 	\hline
  \end{tabular}
  \label{tab:mmaInsts}
  \vspace{-0.25cm}
\end{table}


MMA can operate with several data types that have different sizes.
Moreover, a single MMA instruction can accumulate multiple outer products depending on the size of the data elements.
For instance, for a 32-bit data type, each 128-bit VSR contains four elements and the MMA instruction computes and accumulates a single outer product into the 512-bit accumulator that contains $4 \times 4$ matrix elements~\cite{BhatTR20}.
Following BLAS terminology, such an instruction is called a rank 1 update~\cite{GeijnGoto11}.
However, for a 16-bit data type, each VSR contains eight elements and the MMA instruction computes and accumulates two outer products into the $4 \times 4$-element accumulator, thus computing a rank 2 update.
For 8-bit data types the instruction computes rank 4 updates and for 4-bit data types it computes rank 8 updates.
\rtab{mmaInsts} shows the types supported by MMA.
The computation size indicates the size of each operand and the rank of the update computed by an MMA instruction.
For all types with up to 32 bits the result in the accumulator is 16 32-bit values organized in a $4 \times 4$ grid.
For \code{f64} the accumulator contains $4 \times 2$ elements of the matrix and performs a rank 1 update.

\section{Code Generation for \gemm}
\label{sec:CodeGen}

\ralg{macrokernel} provides an overview of the computation of \gemm within our compiler pass.
The code generation can be divided into two abstract levels: \textbf{macro:} target-independent blocking and packing of matrices for faster memory access and \textbf{micro:} small target-dependent kernel lowered from compiler-intrinsic calls.
A compiler-intrinsic-based  micro kernel, instead of a hand-crafted assembly micro kernel, enables the micro-kernel code to be (i) automatically optimized --- for instance via vectorization --- and mapped --- via instruction selection --- to efficient instructions in the target or (ii) generated by a lowering algorithm that aims to exploit target-specific instructions (See~\rsec{microLevel}).
The macro-level strategy is inspired by the memory-hierarchy modelling described  by Goto and Geijn~\cite{goto2008}.
Nevertheless, our compiler-only macro-level algorithm differs from Goto and Geijn's seminal work on key aspects that better capture modern CPU's and accelerator's features (See \rsec{macro-level-algorithm}).
Implementing this strategy fully inside the compiler has advantages:
\begin{inparaenum}
\item compile-time known features of the target architecture --- e.g. minimum vector-register length --- guide the automatic generation of tiling \& packing code, contrasting with hand-crafted and target-specific implementations in BLAS libraries;
\item a flexible packing layout enables the use of either row or column-major tiles to match the access order in the micro kernel;
\item defining tile sizes for more levels or for different cache/memory organizations only requires changes in the heuristic code itself --- the remaining algorithm code remains untouched; and
\item the packing code --- generated by the compiler instead of hardcoded as in libraries --- can be retargeted to accelerators with, for instance, explicit cache/memory management (e.g. software-managed caches or scratch-pad memories).
\end{inparaenum}

\begin{algorithm}[t]
  \caption{Algorithm overview for \gemm}
  \label{alg:macrokernel}
  \begin{algorithmic}[1]
    \For{j~$\gets$~0, N, step nc}\label{loop1}
      \For{k~$\gets$~0, K, step kc}\label{loop2}
        \State pack(B, BPack, k, j, kc, nc, kr, nr, "B", "Row")\label{packB}
        \For{i~$\gets$~0, M, step mc}\label{loop3}
          \State pack(A, APack, i, k, mc, kc, mr, kr, "A", "Col")\label{packA}
          \For{jj~$\gets$~0, nc step nr}\label{loop5}
            \For{ii~$\gets$~0, mc, step mr}\label{loop4}
              \State AccTile $\gets 0$
              \For{kk~$\gets$~0, kc, step kr}\label{loop6}
                \State BTile~$\gets$~loadTile(BPack, kk, jj, kr, nr, ldb)\label{loadB}
                \State ATile~$\gets$~loadTile(APack, ii, kk, mr, kr, lda)\label{loadA}
                \State ABTile~$\gets$~llvm.matrix.multiply(ATile, BTile, mr, kr, nr)\label{multiply}
                \State AccTile $\gets$ ABTile + AccTile \label{reduction}
              \EndFor
              \State CTile~$\gets$~loadTile(C, i + ii, j + jj, mr, nr, ldc)\label{loadC}
              \If{k == 0}
                \State CTile~$\gets \beta \times$CTile \label{scalebeta}
              \EndIf
              \State CNewTile~$\gets \alpha \times$AccTile \label{scalealpha}
              \State CTile~$\gets$~CTile + NewCTile\label{accumulate}
              \State storeTile(CTile, C, i + ii, j + jj, mr, nr, ldc)\label{storeC}
            \EndFor
          \EndFor
        \EndFor
      \EndFor
    \EndFor
  \end{algorithmic}
\end{algorithm}

In \ralg{macrokernel}, \code{j}, \code{k}, and \code{i} are the offsets of the blocks in the matrices, counted in terms of elements.
The values \code{lda}, \code{ldb} and \code{ldc} are the leading dimensions, and thus the access stride, of the matrices as they are originally stored in memory.
The blocking parameters \mc, \kc, and \nc divide the input matrices \code{A} and \code{B} into blocks properly sized for cache.
The \code{pack} functions in lines~\ref{packB} and~\ref{packA} each create a buffer in main memory containing one copy of an entire block of matrix \code{B} or of matrix \code{A}.
The last argument of the function \code{pack} specifies the data layout within each tile.
The storage order of the tiles after packing, which is independent of the original storage order of the elements of \code{A} and \code{B}, is selected to benefit the tiled multiplication (see \rsec{macro-level-algorithm}).

The tiling parameters \mr, \kr, and \nr ensure optimal resource utilization in the micro kernel.
\code{ATile}, \code{BTile}, \code{AccTile}, \code{ABTile}, \code{CTile}, and \code{CNewTile}, are virtual LLVM IR vectors.
These vectors are allocated and loaded to physical registers by subsequent code-generation passes.
Lines~\ref{loadB} and~\ref{loadA} each load a single tile of an input matrix into an LLVM IR vector from the packed buffers.
The algorithm invokes the intrinsic on Line~\ref{multiply} which multiples \code{ATile} and \code{BTile} and results in a new tile (\code{ABTile}) that is accumulated into \code{AccTile}.
\code{AccTile} is kept in vector regiters for all iterations of the loop on Line~\ref{loop6}.
On the MMA code generation path, \code{AccTile} is mapped to accumulator registers.
A tile of matrix \code{C} is loaded into \code{CTile}, in line~\ref{loadC}, and scaled by the constant $\beta$ on the first iteration of the loop on Line~\ref{loop2}, satisfying GEMM's requirement that the matrix multiplication updates the values of the destination matrix.
Likewise, the accumulation produced in \code{AccTile} is multiplied by the constant $\alpha$ in Line~\ref{scalealpha} satisfy the GEMM mathematical expression.
Once the entire multiplication is completed, \code{CTile} can be stored to its position in memory (Line~\ref{storeC}).

\begin{figure}[t]
  \begin{center}
    \includegraphics[width=1.0\columnwidth]{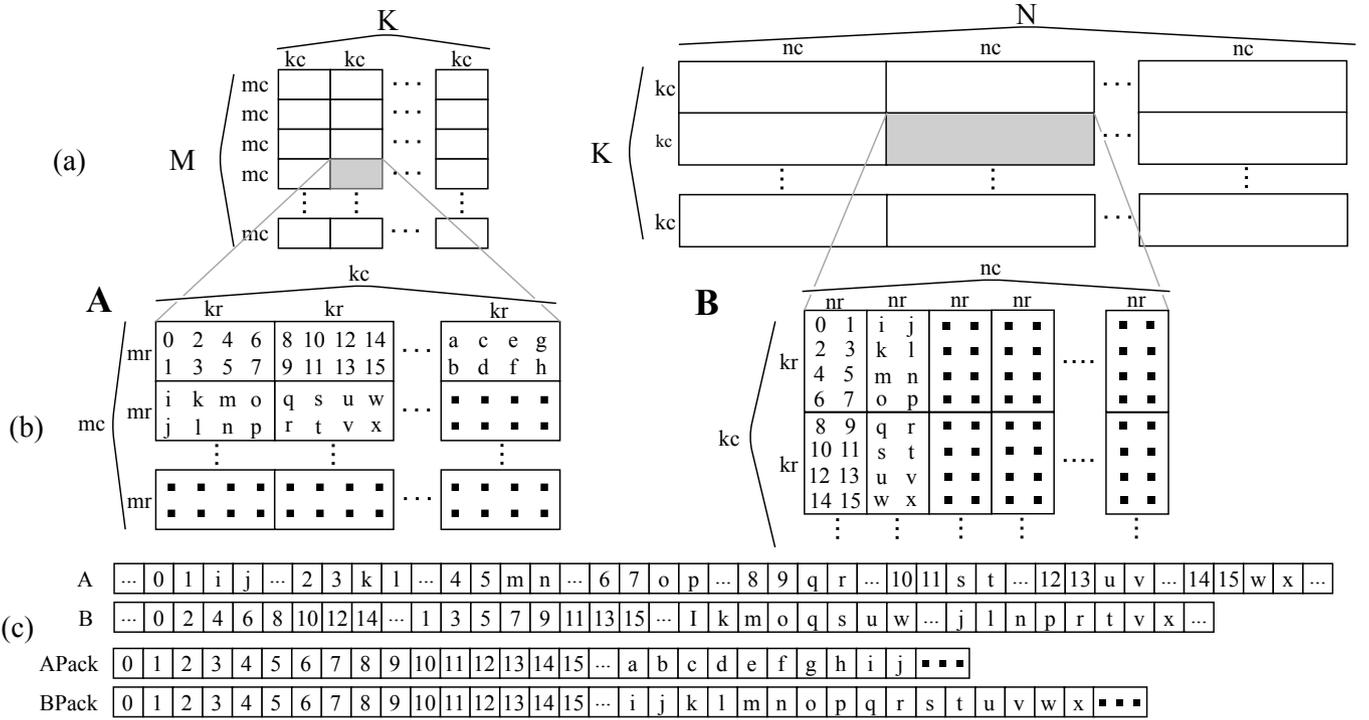}
  \end{center}
  \caption{Tiling and packing for \code{llvm.matrix.multiply}.}
  \label{fig:packing}
  \vspace{-0.25cm}
\end{figure}

\subsection{Macro-level Algorithm: blocking, tiling and packing}
\label{sec:macro-level-algorithm}

Assuming large matrices, portions of each matrix must be brought, through the memory hierarchy, to the registers.
These portions become operands to a mulitiplication intrinsic.
At the micro level, a code generation pass lowers the intrinsic to specific hardware instructions (\rsec{microLevel}).
Contrary to other approaches, no manual vectorization or hand-written assembly code is required.
The whole micro kernel development happens inside the compiler as the intrinsic takes advantage of the target's available matrix engine operations.

In the example shown in \rfig{packing}~(a), matrix \code{A} has $M \times K$ elements and matrix \code{B} has $K \times M$ elements.
Both matrices are stored in memory in column-major order.
A block of matrix \code{A} has $\mc \times \kc$ elements and a block of matrix \code{B} has $\kc \times \nc$ elements.
Each block of \code{A} is divided into tiles of $\mr \times \kr$ elements and each block of \code{B} is divided into $\kr \times \nr$ tiles.
\rfig{packing}~(b) presents the order of the elements and tiles in the packed block when $\mr = 2$, $\kr = 4$ and $\nr = 2$.
These small values are used to keep the figure at a reasonable size.
The actual values for \mr, \kr, and \nr are selected at compilation time to result in a performant micro-level computation for each data-type size in each architecture.

\rfig{packing}~(b) illustrates the partition of a block into tiles.
The tiles are packed within the block in the order in which they will be accessed in the innermost tiling loop.
That is, disregarding the preferred layout of elements within each tile for a particular architecture, the tiles will be placed in rows in the block \mc $\times$ \kc of \code{A} and in columns in the block \kc $\times$ \nc of \code{B}.
The sequential numbers inside the tiles are used simply to indicate the order in which the elements will be accessed when micro-level tiles are loaded into vector registers.
Letters are used at the end of the first row (column) and at the start of the second row (column) of tiles to indicate the relative order of these elements.
\rfig{packing}~(c) shows the layout of the elements of each matrix as they are originally stored in column-major order in memory, and the order of the elements of \code{A} and \code{B} as stored into the \code{APack} and \code{BPack} buffers after packing.
The layout of elements within the tiles is tailored to the needs of the underlying architecture.
For example, \Arm ME, expects a row-major \code{A} and a column-major \code{B} for its input operands and accumulates the result in a row-major C~\cite{ArmISA}.
\rfig{packing}~(c) presents an example for \IBM \PowerTen MMA, which uses column (\code{A}), row (\code{B}), row (\code{C}) layouts.

\begin{eqnarray}
  \kc &\leq& \mathit{L1SizeInBytes} / 2 / \mathit{TypeSizeInBytes} / \VL \label{eq:kc} \\
  \kl &\leq& (\mathit{L1SizeInBytes} / 2 / \mathit{TypeSizeInBytes} - \VL \times \VL) / (2 \times \VL) \label{eq:kl} \\
  \mc &\leq& (\mathit{L2SizeInBytes} - \mathit{L1SizeInBytes}) / \mathit{TypeSizeInBytes} / \kl \label{eq:mc} \\
  \nc &\leq& (\mathit{L3SizeInBytes} - \mathit{L2SizeInBytes}) / \mathit{TypeSizeInBytes} / \kl \label{eq:nc} \\
  &&\kc~\mathit{mod}_{2} \kr = 0 \label{eq:kr} \\
  &&\mc~\mathit{mod}_{2} \mr = 0 \label{eq:mr} \\
  &&\nc \mathit{mod}_{2} \nr = 0 \label{eq:nr}
\end{eqnarray}

Constraints \ref{eq:kc}-\ref{eq:nr} model how the tiling \& packing factors are computed by the compiler-only macro-level algorithm.
\texttt{LXSizeInBytes} is the number of bytes in cache level \texttt{X}, \texttt{TypeSizeInBytes} is the number of bytes in the matrix operations (e.g 4 for single-precision floating-point numbers), and \VL is the minimum vector length size on the target architecture (e.g 4 for 128-bit vector registers).
Similar to Goto \& Van Geijn~\cite{goto2008}, the macro-level algorithm allocates a significant portion of L1 for a piece of each $\kc \times \nc$ block of \code{B}.
However, different from Goto \& Van Geijn, which allocate half of L1 for a $\kc \times \nr$ piece, the macro-level algorithm allocates half of L1 only for a $\kc \times \VL$ piece of \code{B}'s $\kc \times \nc$ block (Constraint~\ref{eq:kc}).
This strategy produces a larger value for \kc to exploit the fact that most modern architecture have enough vector registers to hold $\nr$ multiples of \VL.
The algorithm considers that the remaining half of the L1 will be used to hold $\VL \times \VL$ C elements from memory and $\VL$ elements of \code{A} and \code{B} as per Constraint~\ref{eq:kl}.
\kl is used to maximize the use of L2 for a $\mc \times \kl$ piece of an $\mc \times \kc$ block of \code{A} and to determine how much to allocate in L3 for the $\kl \times \nc$ piece of $\kc \times \nc$ block of \code{B}.
Constaints~\ref{eq:mc} and~\ref{eq:nc} consider the effective size of L2 and L3, respectively, and take into account cache inclusion as in most modern architectures.
The inclusion property of caches is not explicitly modelled by Goto \& Van Geijn.
A final constraint is to make \kc, \mc, and \nc multiple of their respective register-tiling factor in the micro kernel, \kr, \mr, \nr (Constraints~\ref{eq:kr}-\ref{eq:nr}).
\PowerTen's MMA features eight 512-bit accumulators and thirty two 128-bit vector registers to support outer-product computations.
Therefore, for 32-bit matrix elements the performant choice is $\mr = 8$ and $\nr = 16$, as explained in Section~\ref{sec:microLevel}, while \kr is selected to maximize the number of in-accumulator operations.
The effective sizes for L2 and L3 caches may depend on the machine load.
For instance, L3 caches can be shared on some configurations of \PowerTen (\rsec{machine-setup}), and this allows a single-threaded computation to use all of the core's caches.
To account for this effect, the macro algorithm has command line options to provide the effective L2 and L3 cache sizes available per core.

Parameters \mr, \kr, and \nr define the size of the \gemm computed by the micro kernel at each innermost loop iteration in \ralg{macrokernel}.
The values of these parameters are selected by the intrinsic implementation, based on the architectural design of the target.
Section~\ref{sec:microLevel} provides an explanation of how optimal \mr, \kr, and \nr need to be chosen for the micro kernel on \PowerTen MMA.
The values of \mr, \kr, and \nr are the only parameters that need tuning as the cache blocking parameters \mc, \kc and \nc are chosen based on target-specific cache size information available in LLVM.
Following the blocking strategy described by Goto and Geijn, the value of $\kc$ is selected in such a way that an entire row of tiles of the matrix \code{A} --- $\mr \times \kc$ elements --- and an entire column of tiles of the matrix \code{B} --- $\kc \times \nr$ elements --- fit simultaneously in the L1 data cache.
The L1 cache must also have space for an $\mr \times \nr$ tile of the matrix \code{C} but such a tile is quite small in comparison with the space needed for tiles of \code{A} and \code{B}.


\rfig{packing}~(b) shows the order of the matrix elements in the buffers after packing, while~\rfig{packing}~(a) show the original matrices.
Arrays \code{A} and \code{B} in \rfig{packing}~(c) reflect that the original matrices are stored in column-major layout.
Buffers \code{APack} and \code{BPack} in \rfig{packing}~(c) illustrate that, after packing, the elements of matrices \code{A} and \code{B} are stored into these buffers in the order in which they will be accessed by the micro-level computation.
Packing provides two benefits.
First, the tiles of matrices \code{A} and \code{B} lie in the copied memory in the order they will be loaded for the multiply intrinsic in lines~\ref{loadB} and \ref{loadA}.
Second, each row of tiles of \code{A} residing in a part of L1 cache is used in $\lceil  \frac{nc}{nr} \times \frac{kc}{kr}\rceil$ multiplications ($\lceil  \frac{mc}{mr} \times \frac{kc}{kr}\rceil$ times for a column of tiles of \code{B}).

When \code{CNewTile} is not a multiple of a micro tile, the remainder elements are filled with zeroes in the packing buffers and the result is stored with scalar store instructions instead of the efficient matrix store intrinsic used for full tiles.
This brings an overhead for using slow, element-by-element stores instead of fast strided stores.
For the remaining and zero-padded elements, the micro kernel still performs a full computation.


\subsection{Micro-Level Algorithm}
\label{sec:microLevel}

The micro-level algorithm is centered around the LLVM intrinsic \code{llvm.matrix.multiply}.
Intrinsics encapsulate a computational idiom and enable specialization to specific architectures, thus preventing the duplication of code within a compiler infrastructure.
An intrinsic can be viewed as a function call that is eventually replaced by inlined generated code.
Like a function call, an intrinsic has input parameters and a return value.
An intrinsic can be transparently lowered to target-agnostic or to target-specific code.
As expected, the target-agnostic code has lower performance but is available for all LLVM-supported backends.
This section describes the design process to create a target-specific lowering pass for the \code{llvm.matrix.multiply} intrinsic that utilizes the MMA instructions to deliver high performance matrix multiplication.

\begin{lstlisting}[caption={An example usage of \code{llvm.matrix.multiply}.},
      label=lst:intrinsic,numbers=none,language=llvm,float]
%CNewTile = call <128 x float>
  @llvm.matrix.multiply.v128f32.v40f32.
  v80f32(%ATile, %BTile, 8, 5, 16)
\end{lstlisting}

The LLVM IR fragment in \rlst{intrinsic} shows how the \code{llvm.matrix.multiply} intrinsic --- called in line~\ref{multiply} of \ralg{macrokernel} --- takes \code{ATile} and \code{BTile} to produce \code{CNewTile}.
The mangled function name shows the types of \code{\%ATile}, \code{\%BTile}, and \code{\%CNewTile}.
The return value, \code{\%CNewTile} is \code{<128 x float>}, an $8 \times 16$ result, while \code{\%ATile} is \code{<40 x float>}, an $8 \times 5$ tile, and \code{\%BTile} is \code{<80 x float>}, a $5 \times 16$ tile.
The three remaining parameters represent the dimensions of the matrices, $\mr=8$, $\kr=5$, and $\nr=16$, indicating that this intrinsic computes a $C_{8 \times 16} = A_{8 \times 5} \cdot  B_{5 \times 16}$ multiplication.
These tile sizes must be known at compilation time and must match the dimensions of the packed input and output vectors.
The requirement that the tile sizes are known at compile time allows both the target-agnostic lowering and the MMA target-specific lowering to completely unroll the matrix-multiplication loop nest in the IR to give the backend full control over instruction rescheduling.

In addition to these software constraints, scheduling decisions in the algorithm that lowers the intrinsic computation to execute in an MMA backend must also adhere to the following hardware constraints:\begin{inparaenum}
  \item[\fbox{1}] there are at most eight accumulators available per thread and for each accumulator that is used, the usage of four VSRs are blocked;
  \item[\fbox{2}] there are 64 VSRs, thus if eight accumulators are used, there are 32 VSRs remaining to contain the data from the input matrices;
  \item[\fbox{3}] two multiply-and-accumulate outer-product instructions can be issued on a single cycle;
  \item[\fbox{4}] the issue-to-issue latency for the same accumulator is four cycles; and
  \item[\fbox{5}] spilling an accumulator to memory is an expensive operation because it requires an instruction to disassemble the accumulator into four VSRs, four vector store instructions and, later, four vector load instructions.
\end{inparaenum}

\begin{figure}[t]
  \centering
  \begin{tikzpicture}[scale=1/3]
    \foreach \x/\y/\dx/\dy/\c/\o in
        { 0/0/1/4/black!40!white/1,1/0/1/4/black!40!white/1,2/0/1/4/black!40!white/1, 
          3/0/1/4/black!40!white/1,4/0/1/4/black!40!white/1, 
          0/4/1/4/black!70!white/1,1/4/1/4/black!70!white/1,2/4/1/4/black!70!white/1, 
          3/4/1/4/black!70!white/1,4/4/1/4/black!70!white/1, 
          6/9/4/1/black!70!white/1,6/10/4/1/black!70!white/1,6/11/4/1/black!70!white/1, 
          6/12/4/1/black!70!white/1,6/13/4/1/black!70!white/1, 
          10/9/4/1/white/0,10/10/4/1/white/0,10/11/4/1/white/0, 
          10/12/4/1/white/0,10/13/4/1/white/0, 
          14/9/4/1/black!40!white/1,14/10/4/1/black!40!white/1,14/11/4/1/black!40!white/1, 
          14/12/4/1/black!40!white/1,14/13/4/1/black!40!white/1, 
          18/9/4/1/white/0,18/10/4/1/white/0,18/11/4/1/white/0, 
          18/12/4/1/white/0,18/13/4/1/white/0, 
          6/4/4/4/black!70!white/1,10/4/4/4/white/1,14/4/4/4/white/1,18/4/4/4/white/1, 
          6/0/4/4/white/1,10/0/4/4/white/1,14/0/4/4/black!25!white/1,18/0/4/4/white/1} 
    {
      \draw[step=1, shift={(\x, \y)}] (0, 0) grid +(\dx, \dy);
      \ifnum\o=1
        \draw[line width=2,dotted,shift={(\x, \y)}] (0, 0) -- (\dx, 0) -- (\dx, \dy) -- (0, \dy) -- cycle;
      \fi
      \pgfmathsetmacro{\yMax}{\dy-1}
      \pgfmathsetmacro{\xMax}{\dx-1}
      \foreach \yIdx in {0,...,\yMax} {
        \foreach \xIdx in {0,...,\xMax} {
          \pgfmathsetmacro{\xc}{\x+\xIdx}
          \pgfmathsetmacro{\yc}{\y+\yIdx}
          \draw[fill=\c] (\xc, \yc) -- +(1, 0) -- +(1, 1) -- +(0, 1) -- cycle;
        }
      }
    }

    \draw[->,line width=1.25] (-0.75, 6) -- node[left=-.75] {v} +(0, -4);
    \draw[->,line width=1.25] (10, 14.6) -- node[above=-.75] {h} +(8, 0);
    \draw[->,line width=1.25] (1.25, -2) -- node[below=-.75] {k} +(2.5, 0);
    \draw[->,line width=1.25] (24.5, 12.75) -- node[right=-.75] {k} +(0, -2.5);

    \node[above] at (0, 8) {ATile};
    \node[left] at (6, 14) {BTile};
    \node[below] at (22, -0.25) {CNewTile};

    \draw[decorate, decoration={brace, mirror}] (0, -0.3) -- node[below=0.5] {\kr} +(5, 0);
    \draw[decorate, decoration={brace, mirror}] (6, -0.3) -- node[below=0.5] {\nr} +(16, 0);
    \draw[decorate, decoration={brace, mirror}] (22.3, 0) -- node[right] {\mr} +(0, 8);
    \draw[decorate, decoration={brace, mirror}] (22.3, 9) -- node[right] {\kr} +(0, 5);

    \foreach \y in {0, 4} {
    }
    \foreach \x in {6, 10, 14, 18} {
    }

    \foreach \x in {0, ..., 3} {
      \foreach \y in {0, 1} {
        \pgfmathsetmacro{\nx}{(\x + 2) * 4}
        \pgfmathsetmacro{\ny}{(1 - \y) * 4 + 2}
        \pgfmathsetmacro{\i}{int(\x + \y * 4)}
        \node[circle,draw=black,fill=white,inner sep=0, minimum size=15] at (\nx, \ny) {\i};
      }
    }
  \end{tikzpicture}
  \caption{Division of \code{CNewTile} into MMA accumulators.}
  \label{fig:innerTiling}
  \vspace{-0.25cm}
\end{figure}
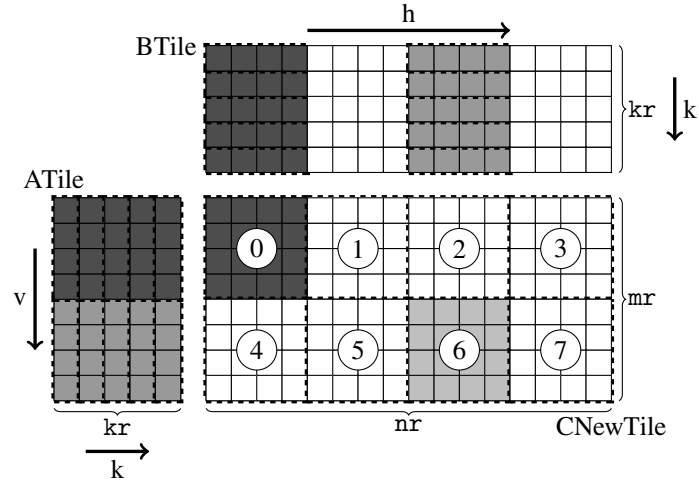

\rfig{innerTiling} illustrates how \code{CNewTile} is divided into portions that are assigned to the MMA accumulators.
\code{ATile}, \code{BTile}, and \code{CNewTile} are represented in two dimensions to illustrate the position of the elements in the matrices.
Each small square in the figure represents one 32-bit element of a matrix.
A circled number indicates that the corresponding portion of \code{CNewTile} is assigned to that accumulator number.
When the intrinsic is executed each accumulator computes \kr outer products using a multiply-and-add operation.
The two tones of gray colour in \rfig{innerTiling} illustrate that a strip of \code{ATile} and a strip of \code{BTile} are used for the accumulation of each portion of \code{CNewTile}.
Each strip is reused for all the accumulations in the same row or column of accumulators.
Each outer-product computation needs two four-element operands, one from \code{ATile} and one from \code{BTile}.
These operands are surrounded by dashed lines for the two accumulations highlighted in gray.
The arrows indicate how the loop indices in \ralg{intrinsic} iterate for the example in the figure.

\begin{algorithm}[t]
  \caption{Computation by \code{llvm.matrix.multiply}}
  \label{alg:intrinsic}
  \begin{algorithmic}[1]
    \Function{\code{llvm.matrix.multiply}}{ATile, BTile, nr, kr, mr}
    \State CNewTile~$\gets$~Empty\label{createC}
    \State Assemble ACCs and initialize to zero\label{zeroAccs}
    \For{$\text{k}=0$ \textbf{to} kr-1}\label{extractLoopStart}
      \For{$\text{v}=0$ to VAccs-1}\label{VextractLoop}
        \State AOps[v][k]~$\gets$~Extract op from ATile[v][k]\label{aOps}
      \EndFor
      \For{$\text{h}=0$ to HAccs-1}\label{HextractLoop}
         \State BOps[h][k]~$\gets$~Extract op from BTile[h][k]\label{bOps}
      \EndFor
    \EndFor
    \For{$\text{k}=0$ \textbf{to} kr-1}\label{accLoopStart}
      \For{$\text{v}=0$ to VAccs-1}\label{tileLoop1}
        \For{$\text{h}=0$ to HAccs-1}\label{tileLoop2}
          \State Accs[v][h] += \label{builtinUse}
          \Statex[5] MMABuiltIn(AOps[v][k], BOps[h][k])
        \EndFor
      \EndFor
    \EndFor\label{accLoopEnd}
    \State Disassemble ACCs and store VSRs into CNewTile\label{disStore}
    \State \Return CNewTile\label{return}
    \EndFunction
  \end{algorithmic}
\end{algorithm}

\ralg{intrinsic} describes the lowering of the intrinsic computation for MMA.
The compile-time constants \code{VAccs} and \code{HAccs} (used in lines~\ref{VextractLoop} and \ref{HextractLoop}) specify the layout of the accumulators for the computation.
For the example in \rfig{innerTiling}, \code{VAccs = 2} and \code{HAccs = 4}.
These constants in the compiler generalize the lowering and make it applicable to future architectures where the ideal arrangement to increase data reuse may be different from the $2 \times 4$ arrangement in the \PowerTen processor.

After creating \code{CNewTile} (line~\ref{createC}) and assembling and zeroing all the accumulators (line~\ref{zeroAccs}), \ralg{intrinsic} iterates \code{k} from $0$ to $\kr-1$ (line~\ref{extractLoopStart}) to extract operands from \code{ATile} and \code{BTile} into virtual IR registers.
For each value of k, using the accumulator assignment shown in \rfig{innerTiling}, the algorithm extracts two operands from \code{ATile} (lines~\ref{VextractLoop}-\ref{aOps}) and four operands from \code{BTile} (lines~\ref{HextractLoop}-\ref{bOps}).
For $k=0$ the operands are extracted from the leftmost column of \code{ATile} and from the top row of \code{BTile} in \rfig{innerTiling}.
The algorithmic presentation in \ralg{intrinsic} uses the notation \code{AOps[v][k]} and \code{BOps[h][k]} to show the connection between the operands extracted from \code{ATile} and \code{BTile} with the use of these operands in the \code{MMABuiltin} on line~\ref{builtinUse}.
In the compiler, at this point in the lowering, each four-element operand is extracted into a virtual IR register.
The actual VSRs used for each operand are determined later by a register-allocation pass.

In \rfig{innerTiling} each operand is formed by four elements and, once extracted, occupies one 128-bit VSR.
Given constraints \fbox{1} and \fbox{2}, with the choice of $\kr = 5$, there are enough non-blocked VSRs to contain all the thirty operands needed for the computation illustrated in \rfig{innerTiling}.
Thus, laying out the accumulators in this $2 \times 4$ pattern maximizes the reuse of values loaded into the VSRs: operands extracted from \code{ATile} are reused four times and operands extracted from \code{BTile} are reused two times.

Once all operands are extracted into VSRs, the algorithm again iterates over the dimension \kr to compute each piece of \code{C} to avoid spilling any accumulator to memory.
Following constraint \fbox{3}, two outer-product instructions are issued in each cycle.
Four pairs of accumulators can be scheduled before circling back to the first pair, thus satisfying constraint \fbox{4}.
The assignment of a portion of \code{CNewTile} to a single accumulator eliminates the need to spill accumulators, thus increasing the performance according to constraint \fbox{5}.

The lowering of the intrinsic for execution in \PowerTen is based on a set of builtins that encapsulate the computation of an outer product.
There is a set of builtins for each data type to allow the code generator to select a multiply-and-add that either initializes or updates an accumulator.
All combinations of positive/negative multiplication with positive/negative accumulation are available as well.
For some data types there are also builtins that perform saturating arithmetic instead of overflow for accumulation.
Thus, when lowering the intrinsic for the GEMM computation, the compiler selects the appropriate positive multiply and positive accumulate builtin for the specified data type which is then used on line~\ref{builtinUse}.

\subsection{Other Data Types}
The presentation so far assumed 32-bit data types where each operand VSR contains 4 elements and an MMA instruction computes a rank 1 update, computing and accumulating a single outer product.
Halving the data-type size doubles the number of elements in each VSR and doubles the rank of the update.
For example, for a 16-bit data type MMA computes a rank 2 update while an 8-bit data type computes a rank 4 update.
The packing of more elements into a single VSR and the accumulation of multiple outer products by a single MMA instruction requires changes to \ralg{macrokernel} and \ralg{intrinsic}.
Let $n$ be the number of outer products performed by an MMA instruction --- i.e. the rank of the update.
Now the step size of the loops on lines~\ref{extractLoopStart} and \ref{accLoopStart} must both be $n$ because, in \rfig{innerTiling}, $n$ rows of \code{BTile} and $n$ columns of \code{ATile} are packed into each VSR.
The extraction of operands in lines~\ref{aOps} and \ref{bOps} is now a strided access.
For instance, for $n=2$ (16-bit data types), four consecutive elements are extracted from row $k$ and four consecutive elements are extracted from row $k+1$ to form the 128-bit VSR.
The length of \kr must increase by $n$ times to provide enough data to populate the VSRs.
The effect is that more partial-product accumulations can be computed per micro-kernel invocation given the same number of assemblies and disassemblies because the number of multiplications per outer product increases by $n$.

For double-precision floating-point data type \code{f64} an accumulator contains $4 \times 2$ 64-bit elements.
The operand extracted from \code{ATile} is placed into a combination of two VSRs that together contain four elements while the operand extracted from \code{BTile} is placed into a single 128-bit VSR containing two elements.
Therefore, for \code{f64} the value of \nr should be reduced in half to reflect the number of VSRs available.
With this reduction, an \code{ATile} tile occupies 16 VSRs and a \code{BTile} tile also occupies 16 VSRs.
The extraction of operands into vector registers in lines~\ref{aOps} and~\ref{bOps} of \ralg{intrinsic} must be changed accordingly.

\subsection{Arbitrary Values for \nr, \mr, \kr and Access Order}
Until now, the algorithms have used values of \mr and \nr selected such that a micro kernel with the accumulator arrangement shown in \rfig{innerTiling} could be computed with a single set of assemble and disassemble instructions.
However, the implementation of \ralg{intrinsic} in LLVM must handle any \code{llvm.matrix.multiply} intrinsic created by any compilation path and thus must handle arbitrary values for \nr, \mr and \kr.
The code-lowering algorithm also supports inputs and outputs in any access order through modifications to the functions that extract operands and store the results in the accumulators to memory.

To handle larger values of \mr and \nr, the micro-level code-lowering algorithm has an additional outer double-nested loop that logically divides the \code{CVec} tile into $8 \times 16$-element sections as shown in \rfig{innerTiling}.
Each of these sections can then be handled as shown in \ralg{intrinsic}.
The disadvantage of a tile size that spans  multiple accumulator sections is that the extraction of data into vector registers becomes more complex.
For example, consider a 32-bit data multiplication as shown \rfig{innerTiling} but with the values of \nr and \mr double of what is shown in the figure.
The rows of \code{ATile} and \code{BTile} shown in \rfig{innerTiling} are now a portion of the rows of larger tiles and the data extraction must gather the correct data into the vector registers that will be used by the accumulators.
This data gathering adds additional code and may impact access locality if the tiles are large enough.
Moreover, if \ralg{intrinsic} is used in combination with \ralg{macrokernel}, then the packing work done earlier in lines~\ref{packB} and \ref{packA} of that algorithm may not result in optimal locality.
As well, if \kr is smaller than $K$ (see \rfig{packing}) multiple invocations of the intrinsic are needed to compute each element of the result matrix.
Therefore, accumulators must be assembled and disassembled multiple times, creating an issue with constraint \fbox{5}.
\section{A performant Compiler-Only Solution}
\label{sec:results}

Experimental results support the following claims:
\begin{inparaenum}
\item the macro-level algorithm is architecture-independent: it is performant across four different architectures (\Intel's and \AMDREG's x86; \IBM's \PowerNine and \PowerTen);
  \item the algorithm, when fully implemented inside the compiler, surpasses the performance of PluTo, a widely used polyhedral-based compiler-only approach;
  \item the macro-level algorithm, even when coupled with a generic-lowering of the LLVM's matrix-multiply intrinsic, approaches the performance of Eigen~\cite{eigenweb} and OpenBLAS~\cite{2012xianyi};
  \item for small \gemms the compiler-only approach performance can surpass library calls;
  \item the MMA lowering delivers more than $2.6$x the performance of VSX on \PowerTen; and
  \item our compiler-only solution boosts \gemm's performance by exploiting target-specific matrix engines: the micro-level algorithm lowers multiply-add \gemm reductions to efficient IBM's MMA instructions resulting in better performance than Eigen and up to $96\%$ of OpenBLAS' peak performance.
\end{inparaenum}

\subsection{Experimental Setup}
\label{sec:setup}

\subsubsection{Machine Setup}
\label{sec:machine-setup}

\begin{table}[t]
\centering
\caption{Machine configuration used in the evaluation.}
\vspace{-0.25cm}
\addtolength{\tabcolsep}{-2.0pt}
{\footnotesize \begin{tabular} {p{1.6cm}| p{1.48cm} | p{1.25cm} | p{1.35cm} |p{1.6cm}}
  \makecell[l]{Component\\ / Processor}
                        & \Intel~\XeonTM 8268 & \IBM \PowerNine & \IBM \PowerTen & \AMD~\EpycTM 7742  \\\hline
  Cores/Threads         & 24/2              & 20/4            & 8/8            & 64/2              \\
  L1 i-cache            & 32KiB             & 32KiB           & 32KiB          & 32KiB             \\
  L1 data cache         & 32KiB             & 32KiB           & 48KiB          & 32KiB             \\
  L2 (unified)          & 256KiB            & 512KiB          & 1024KiB        & 512KiB            \\
  L3 (unified)          & 35.75MiB          & 10MiB           & 4MiB           & 16MiB             \\
  RAM                   & 755GiB            & 1TiB            & 1TiB           & 995GiB            \\
  Memory bandwidth      & 131.13GiB/s       & 140GiB/s        & --             & 190.7GiB/s        \\
\end{tabular}
}
\label{tab:machines}
\vspace{-0.35cm}
\end{table}

Experimental evaluation used the four platforms shown in \rtab{machines}\footnote{Core counts are per socket and thread counts are per core. Cache sizes in \PowerTen are those available for a single thread.}.
All platforms run Linux with 64-bit kernel at version \code{4.15.0-155-generic}, except \PowerTen which runs at version \code{4.18.0-277.el8.ppc64le}.
L1 and L2 cache sizes  are per core while L3 is shared among all cores on \Intel's and \AMD's machines.
L3 cache is shared between pairs of cores in \PowerNine but is local to a core in \PowerTen.

\subsubsection{Compiler Options}
\label{sec:compiler-flags}

All binaries are compiled with Clang version $14$ at \code{-O3} targeting each architecture.
Binaries for \Intel are targeted and tuned to  \CascadeLake (\code{-march=cascadelake -mtune=cascadelake}), \AMD's to \ZenTwo (\code{-march=znver2 -mtune=znver2}), and \IBM's machines to their corresponding CPUs (\code{-mcpu=\{power9\|power10\} -mtune=\{power9\|power10\}}).
All binaries are statically linked (\code{-static}).

\subsubsection{Code Generation and Libraries}
\label{sec:codegen-and-libraries}

\rsec{results-claim1} and \rsec{results-claim2} show results for the following code generation strategies:

\begin{itemize}
  \item \textbf{\Intrinsic}:
    \gemm loops are replaced with a single call to LLVM's matrix-multiply intrinsic.
    This option causes all \gemm loops to be unrolled and multiply-add computations to be lowered with either generic or MMA lowering (\rsec{microLevel}).
  \item \textbf{\Tiling}:
    This option tiles the \gemm loops iterating over each dimension (M, K, and N) and calls the matrix-multiply intrinsic in the body of the innermost loop (depth 3).
    Tile sizes are computed following Goto et al.'s strategy as described in \rsec{macro-level-algorithm}.
  \item \textbf{\TilingPacking}:
    This option tiles the \gemm loops and packs the input matrices \code{A} and \code{B} as described in \rsec{macro-level-algorithm}.
    The matrix-multiply intrinsic is called in the body of the innermost loop (depth 6) to compute a block of \code{C} from tiles of \code{A} and \code{B}.
  \item \textbf{\PLUTO}:
    This option applies a source-to-source transformation using PLuTo\footnote{PLuTo~\href{https://github.com/bondhugula/pluto}{Release v0.11.4}}, a polyhedral-based paralelism and locality optimizer, that automatically tiles and redorders loops annotated with \code{#pragma scope}.
    PluTo's auto-tiling\footnote{\code{-tile -l2tile}} for both the first and second-level caches are enabled.
  \item \textbf{\BLAS}:
    This option replaces the \gemm loops with a single call to the CBLAS interface \gemm routine.
    The OpenBLAS\footnote{OpenBLAS~\href{https://github.com/xianyi/OpenBLAS}{\code{1e4b2e98d953a18df85243a3fa019a105cbcb3dc}}.}~\cite{2012xianyi} library is the implementation of the CBLAS interface evaluated.
    The library is compiled and linked as described in \rsec{compiler-flags}.
  \item \textbf{\EIGEN}:
    This option replaces the \gemm loops with Eigen\footnote{Eigen~\href{https://gitlab.com/libeigen/eigen}{\code{5bbc9cea93ef29cee2b8ffb2084d4ebca32600ba}}}~\cite{eigenweb} code to compute the general matrix-matrix multiply.
    Eigen code is compiled as described in \rsec{compiler-flags}.
\end{itemize}

The experiments use versions of \BLAS and \EIGEN, which support both VSX and MMA instructions, that were contributed to the open source community by internal teams from \IBM.
All versions listed above are compiled and set to execute single-threaded code.
A naive implementation of a $M \times K \times N$ \sgemm in C++ serves as the base code for our algorithm.
Carvalho et al.'s pattern-matching algorithm~\cite{CarvalhoTACO21} is used to automatically identify and replace these \gemm loops in the source code with the code generated by the macro-level algorithm.
The input matrices are stored and accessed in column-major order and the values $\mr=16$, $\nr=4$, and $\kr=64$ are used for all platforms, except on \PowerTen which used $\mr=16$, $\nr=8$, and $\kr=128$.

Even though \PLUTO lacks critical optimizations, such as packing, it is the only compiler-only solution available that produced correct results for the programs used in the experimental evaluation. 
Polly, another compiler-only solution,  either failed to optimize the input code or generated code that computes incorrect results.\footnote{These incorrect results were reported to the developers of Polly.}

This paper presents an end-to-end solution that compiles a C/C++ input program and replaces identified \gemm loops with the high-performance code generated by the macro-level algorithm.
Another compiler-only solution, introduced by Uday Bondhugula, is not an end-to-end solution  because it relies on a hand-crafted MLIR code written in the Affine dialect~\cite{uday2020}.
At the time of writing, there is no way to automatically translate C/C++ programs to MLIR.
Thus, Bondhugula's hand-crafted input benefits from MLIR passes that are unreachable for a C/C++ end-to-end compilation path.
These MLIR passes are also not reachable to compile \EIGEN's and \BLAS's code, our baselines.
The experimental results in this Section indicate that the macro-level algorithm and the generic-lowered micro kernel produce code that reaches comparable performance to Bondhugula's approach relative to \BLAS~\cite{uday2020}.
Moreover, the MMA lowering reaches up to $96\%$ of \BLAS's peak performance, while Bondhugula's reached only $91\%$ of \BLAS's performance on a Coffee Lake Intel machine with his MLIR code that employes a target-specific vectorization pass~\cite{uday2020}.

\subsubsection{Experimental Methodology}
\label{sec:methodology}

The methodology follows these steps:
\begin{inparaenum}
 \item Compile each benchmark, for each platform, with aggressive optimization flags (see \rsec{compiler-flags}) using the six strategies: \Intrinsic, \Tiling, \TilingPacking, \PLUTO, \BLAS, and \EIGEN.
 \item Create a list containing all of the executables.
 \item Measure the execution time of each element of the list once.
 \item Randomize the list of executables before the next set of measurements.
 \item Repeat until there are twenty measurements for each executable.
\end{inparaenum}
This methodology ensures that changes to the execution environment that may affect performance manifest as a higher variance between executions of the same executable rather than as a bias in the measurements for a version of the experiements.
\rsec{results-claim1} and \rsec{results-claim2} show $95\%$ confidence intervals.

\subsection{Performance Comparison Against Other Compiler-Only Approaches}
\label{sec:pluto-results}

\begin{figure*}[t]
  \centering
  \begin{subfigure}{0.45\linewidth}
    \includegraphics[width=1.0\linewidth]{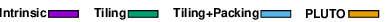}
  \end{subfigure}\\
  \begin{subfigure}{0.31\linewidth}
    \includegraphics[width=1.0\linewidth]{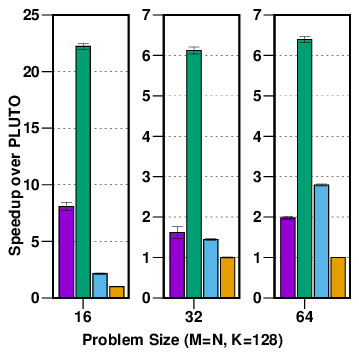}
    \subcaption{\Intel x86}
  \end{subfigure}
  \begin{subfigure}{0.31\linewidth}
    \includegraphics[width=1.0\linewidth]{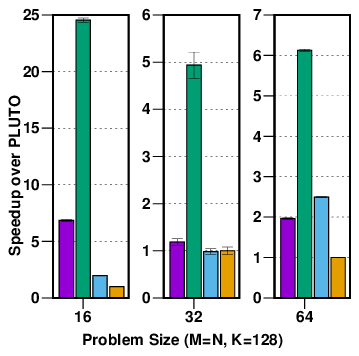}
    \subcaption{\AMD x86}
  \end{subfigure}
  \begin{subfigure}{0.31\linewidth}
    \includegraphics[width=1.0\linewidth]{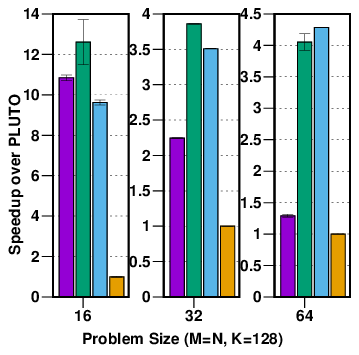}
    \subcaption{\IBM \PowerNine}
  \end{subfigure}
  \caption{Speedup over \PLUTO for small \sgemm in each platform.}
  \label{fig:speedup-pluto-small}
  \vspace{-0.35cm}
\end{figure*}%
\begin{figure*}[t]
  \centering
  \begin{subfigure}{0.35\linewidth}
    \includegraphics[width=1.0\linewidth]{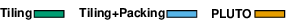}
  \end{subfigure}\\
  \begin{subfigure}{0.31\linewidth}
    \includegraphics[width=1.0\linewidth]{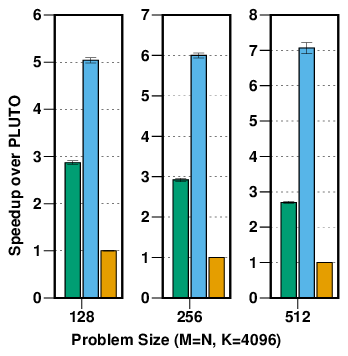}
    \subcaption{\Intel x86}
  \end{subfigure}
  \begin{subfigure}{0.31\linewidth}
    \includegraphics[width=1.0\linewidth]{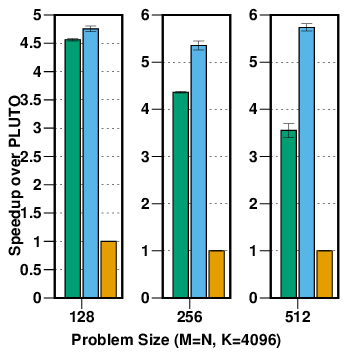}
    \subcaption{\AMD x86}
  \end{subfigure}
  \begin{subfigure}{0.31\linewidth}
    \includegraphics[width=1.0\linewidth]{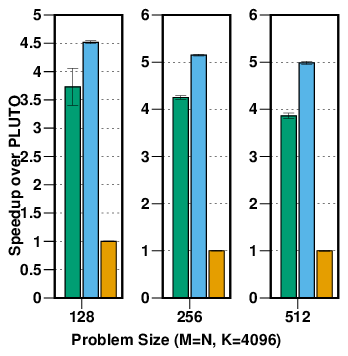}
    \subcaption{\IBM \PowerNine}
  \end{subfigure}
  \caption{Speedup over \PLUTO for medium \sgemm in each platform.}
  \label{fig:speedup-pluto-medium}
  \vspace{-0.35cm}
\end{figure*}%
\begin{figure*}[t]
  \centering
  \begin{subfigure}{0.35\linewidth}
    \includegraphics[width=1.0\linewidth]{images/legend-pluto-all-minus-intrinsic.eps}
  \end{subfigure}\\
  \begin{subfigure}{0.31\linewidth}
    \includegraphics[width=1.0\linewidth]{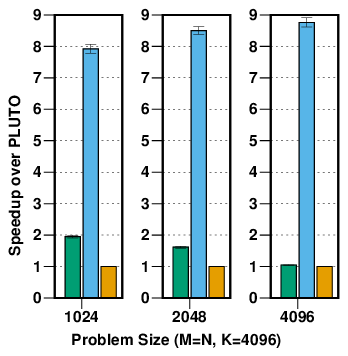}
    \subcaption{\Intel x86}
  \end{subfigure}
  \begin{subfigure}{0.31\linewidth}
    \includegraphics[width=1.0\linewidth]{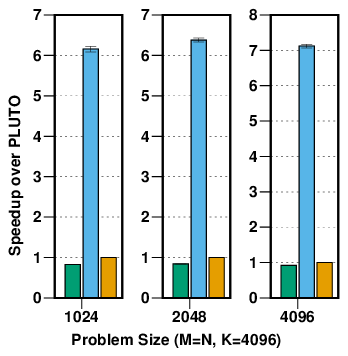}
    \subcaption{\AMD x86}
  \end{subfigure}
  \begin{subfigure}{0.31\linewidth}
    \includegraphics[width=1.0\linewidth]{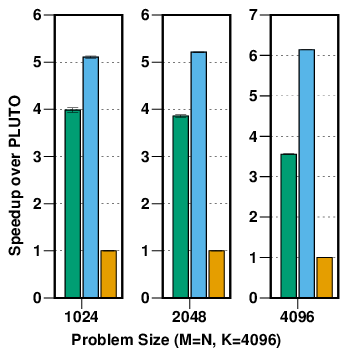}
    \subcaption{\IBM \PowerNine}
  \end{subfigure}
  \caption{Speedup over \PLUTO for large \sgemm in each platform.}
  \label{fig:speedup-pluto-large}
  \vspace{-0.35cm}
\end{figure*}

The results in \rfig{speedup-pluto-small} indicate that for small \sgemm sizes the performance of \Tiling is far superior than \PLUTO across all platforms: \Tiling is up to $22$x faster than \PLUTO for the smallest \sgemm size on \Intel, almost $25$x faster on \AMD, and over $11$x faster on \PowerNine.
For small problem sizes \Tiling performs best overall as the matrices fit well in the memory hierarchy.
Given that small matrices fit in cache,  \TilingPacking only adds overhead due extra memory movent for packing input matrices, performing worse than \Tiling, but still better than \PLUTO.
In addition, \Tiling is aware of the vector unit capacity in the target architecture and generates a micro-kernel that fully utilizes the vector unit.
\PLUTO performs poorly because its auto-tiling mechanism generates innerloops with conservative tiling sizes which do not saturate the vector unit capacity.

The graphs for medium and large \sgemms do not include results for \Intrinsic because the LLVM matrix-multiply intrinsic is designed for small kernels and completely unrolls the loops.
Invoking that intrinsic with large dimensions leads to prohibitive compilation times.

\rfig{speedup-pluto-medium} shows speedup results for medium \sgemm sizes.
Overall, \TilingPacking performs best across all platforms.
Medium-sized matrices still fit in the low-level caches (L2 and L3), however not in L1.
As matrices become larger, they span across multiple memory pages.
Therefore, the data reorganization performed by packing improves the utilization of the memory hiearchy by increasing data temporal locality and reducing both TLB and page-faults.
\PLUTO does not employ packing and thus continues to perform poorly for medium-sized matrices.
On \AMD, \Tiling is still compatitive with \TilingPacking due to the larger caches and lower cache access latency.
Nevertheless, as \rfig{speedup-pluto-large} shows, with large \sgemm sizes that no longer fit the low-level caches (L2 and L3), the performance between \Tiling and \TilingPacking widens.
On \PowerNine, \Tiling does not have its performance degraded as much as on \Intel and \AMD for large problems due to large cache line sizes on \PowerPC.
However, \TilingPacking remains the best strategy for large \sgemms due to higher data temporal locality and significantly smaller TLB and page-faults in contrast to \Tiling.

LLVM has a polyhedral optimizer, Polly~\cite{polly}, that is not enabled by default.
If Polly's optimizations are enabled\footnote{\code{-mllvm -polly}} the code runs more than $4.8\times$ slower than \TilingPacking on all three architectures.
Polly also has an option to parallelize loops\footnote{\code{-mllvm -polly-parallel -lgomp}}.
When this parallelization is run on 20 threads, the resulting code is $1.4\times$ slower than \TilingPacking on \PowerNine.
Unfortunately the \sgemm results computed with Polly-optimized code are incorrect~\footnote{Polly maintainers are aware of this correctness issue.}. 
Thus, the performance results for Polly cannot be included in the comparison at this time.
As reported by Carvalho et al.~\cite{CarvalhoTACO21}, we also failed to reproduce the performance results reported by Gareev et al.'s recent work~\cite{GareevTACO18}.
Enabling Polly's \gemm idiom recognition and optimization pass\footnote{\code{-polly-pattern-matching-based-opts=true}} the code runs $4.8\times$ slower than \TilingPacking.
Therefore, to the best of our knowledge, this work is the first to present a compiler-only solution that produces performance that is comparable to high-performance libraries (\rsec{results-claim1}) without relying on hand-written assembly micro kernels~\cite{uday2020}.

LLVM's default lowering of matrix-load intrinsics generates code that spills data from both input operands loaded for the matrix-multiply intrinsic.
The maintainers of the matrix intrinsic passes provide an alternative lowering for the matrix-multiply intrisic that bypasses matrix-load intrinsics and generates an unrolled sequence of load-load-multiply instructions with smaller load instructions.
This alternative lowering was used as a work-around to eliminate the spill problem on \Intel, \AMD, and \PowerNine.
However, this alternative lowering is currently not modularly integrated into the framework and thus could not be used for the micro-level algorithm.
Therefore, the spill problem for \Intrinsic, \Tiling, and \TilingPacking on \PowerTen was resolved by applying a code transformation after the micro-kernel lowering to sink load instructions -- which load the operands of matrix-multiply -- closer to their uses -- MMA outer-product instructions.

\subsection{Performance Comparison Against High-Performance Libraries}
\label{sec:results-claim1}

\begin{figure*}[t]
  \centering
  \begin{subfigure}{0.45\linewidth}
    \includegraphics[width=1.0\linewidth]{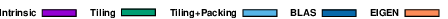}
  \end{subfigure}\\
  \begin{subfigure}{0.32\linewidth}
    \includegraphics[width=1.0\linewidth]{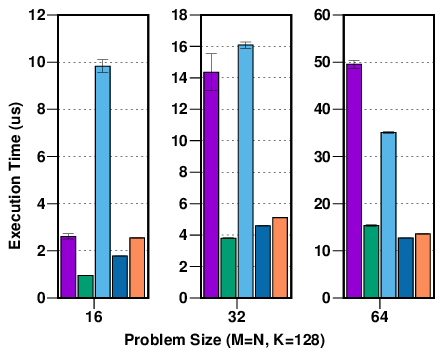}
    \subcaption{\Intel x86}
  \end{subfigure}
  \begin{subfigure}{0.32\linewidth}
    \includegraphics[width=1.0\linewidth]{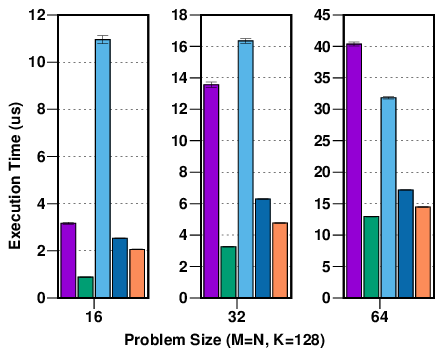}
    \subcaption{\AMD x86}
  \end{subfigure}
  \begin{subfigure}{0.32\linewidth}
    \includegraphics[width=1.0\linewidth]{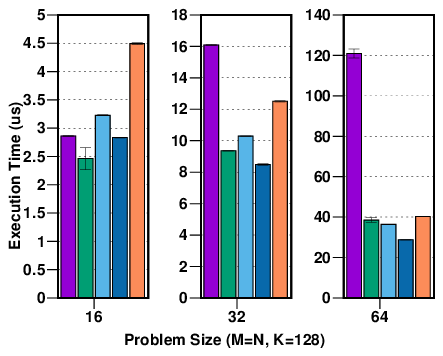}
    \subcaption{\IBM \PowerNine}
  \end{subfigure}
  \caption{Execution time of small \sgemm in each platform.}
  \label{fig:execution-time-libs-small}
  \vspace{-0.35cm}
\end{figure*}%
\begin{figure*}[t]
  \centering
  \begin{subfigure}{0.35\linewidth}
    \includegraphics[width=1.0\linewidth]{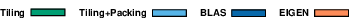}
  \end{subfigure}\\
  \begin{subfigure}{0.32\linewidth}
    \includegraphics[width=1.0\linewidth]{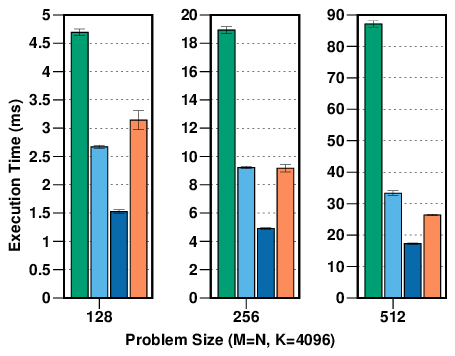}
    \subcaption{\Intel x86}
  \end{subfigure}
  \begin{subfigure}{0.32\linewidth}
    \includegraphics[width=1.0\linewidth]{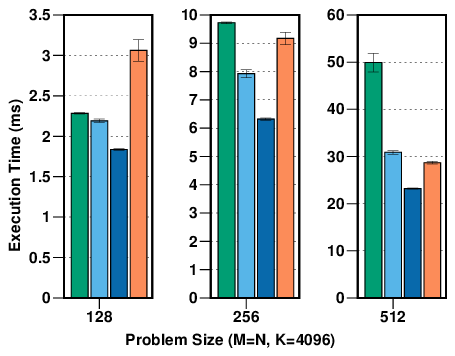}
    \subcaption{\AMD x86}
  \end{subfigure}
  \begin{subfigure}{0.32\linewidth}
    \includegraphics[width=1.0\linewidth]{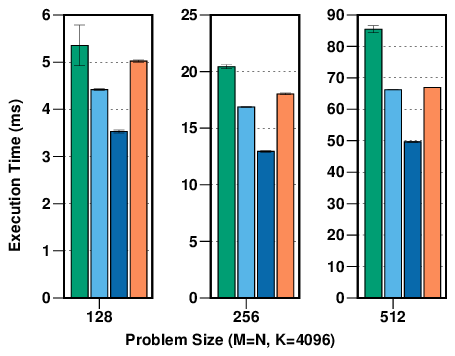}
    \subcaption{\IBM \PowerNine}
  \end{subfigure}
  \caption{Execution time of medium \sgemm in each platform.}
  \label{fig:execution-time-libs-medium}
  \vspace{-0.35cm}
\end{figure*}%
\begin{figure*}[t]
  \centering
  \begin{subfigure}{0.35\linewidth}
    \includegraphics[width=1.0\linewidth]{images/legend-libs-all-minus-intrinsic.eps}
  \end{subfigure}\\
  \begin{subfigure}{0.32\linewidth}
    \includegraphics[width=1.0\linewidth]{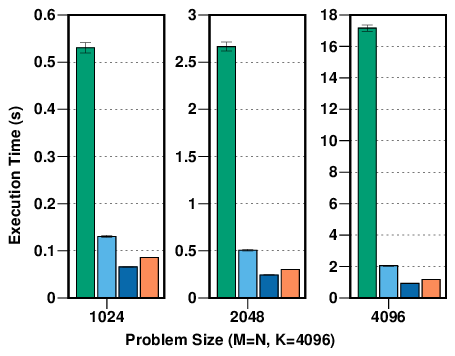}
    \subcaption{\Intel x86}
  \end{subfigure}
  \begin{subfigure}{0.32\linewidth}
    \includegraphics[width=1.0\linewidth]{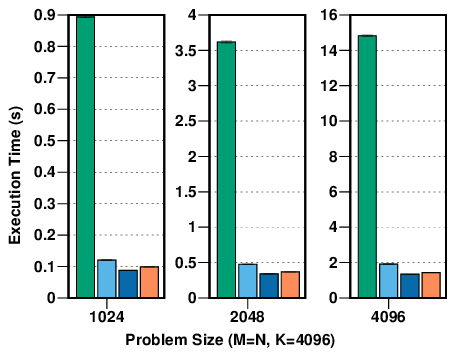}
    \subcaption{\AMD x86}
  \end{subfigure}
  \begin{subfigure}{0.32\linewidth}
    \includegraphics[width=1.0\linewidth]{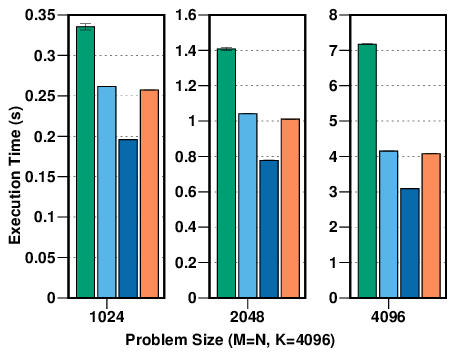}
    \subcaption{\IBM \PowerNine}
  \end{subfigure}
  \caption{Execution time of large \sgemm in each platform.}
  \label{fig:execution-time-libs-large}
  \vspace{-0.35cm}
\end{figure*}

The results in \rfig{execution-time-libs-small} indicate that for the smallest \sgemm size the performance of \Intrinsic is on-par with both \EIGEN and \BLAS, thus revealing that the backend alone generates efficient code for computing very small \sgemms.
The superior performance of \Tiling is due to the better utilization of the vector registers in the target architecture. 
\Tiling is more than $85\%$ faster than \BLAS and over $2.6\times$ faster than \EIGEN on \Intel for the problem size $M=N=16$.
On \AMD, \Tiling is more than $2.3\times$ faster than \EIGEN and $2.8\times$ faster than \BLAS for the same problem size.
\Intrinsic performs worse in comparison with the libraries as the problem size increases because, lacking tiling, it exhibits lower cache locality.
In summary, \rfig{execution-time-libs-small} indicates that the compiler-only macro-level approach can surpass high-performance libraries for small problem sizes on multiple platforms.

For medium problem sizes \TilingPacking is the best code-generation strategy overall and it is on-par with \EIGEN across all platforms for the problem size $M=N=128$.
As the size of matrices increases the performance of \Tiling degrades with respect to both \BLAS and \EIGEN because LLVM's matrix-load intrinsic lowering results in additional spill code.
However, \TilingPacking remains competitive across all platforms by matching \EIGEN's performance and is less than $80\%$ slower than \BLAS for $M=N=512$.
As discussed in \rsec{pluto-results}, \Tiling and \TilingPacking show similar performance due to the larger caches and lower cache access latency on \AMD.

For large matrices \Tiling performs worst overall because it produces a data layout that leads to more cache misses.
\Tiling increases the cache utilization within a given dimension but makes cross-dimension accesses expensive because elements of a given column, but different rows, are located on different virtual-memory pages (e.g. 8KB apart on column-major matrix of 2048 columns).
The packing data reorganization creates a better element order that increases data locality on both dimensions.
Therefore, \TilingPacking performs significantly faster than \Tiling on \Intel and \AMD.
On \PowerNine, the gap between \Tiling and \TilingPacking is not as large as in other platforms because \PowerNine has larger cache lines (128 bytes) -- twice as large as on the other platforms -- which, coupled with prefetching, helps to hide memory latency.
The \Tiling performs significantly worse than \TilingPacking on \PowerNine as the problem size increases due to increased \TLB misses and page-faults.
A similar pattern appears on both Intel and AMD machines, where \Tiling suffers memory penalties as the problem size increases, leaving \TilingPacking as the best performing code-generation strategy overall.

Compiling the largest \sgemm with \code{Clang -O3}\footnote{\code{clang -O3 -mcpu=power9 -mtune=power9 -ffast-math -ffp-contract=fast}}, the state of the art when using \code{Clang} prior to this work, cam result in a code that is more than $68\times$ slower than \BLAS.
This work significantly reduces this performance gap.
For the largest \sgemm size ($M=N=4096$), \TilingPacking is slower than \BLAS: over $2.2\times$ slower on \Intel and $43\%$ slower on \AMD.
In comparison with \EIGEN, \TilingPacking is $1.75\times$ slower on Intel and $34\%$ slower on AMD.
On \PowerNine, \TilingPacking is $34\%$ slower than \BLAS and on-par with \EIGEN.
The performance gap between \BLAS and \TilingPacking is in great part due to a performance gap between the \BLAS micro kernel and the generic-lowered micro kernel available upstream in LLVM.
The generic-lowered micro kernel is $2.1\times$ slower on \Intel, $70\%$ slower on \AMD, and $30\%$ slower on \PowerNine than the micro kernel from \BLAS.

Portability is a key design goal for the macro-level algorithm because it leads to the generation of a micro kernel -- for any target architecture available in LLVM -- with significant performance, as indicated by the results above.
Nevertheless, the generic-lowered micro kernel is not a match for the hand-crafted assembly micro kernel available to high-performance libraries.
This performance gap can be reduced --- as demonstrated for \PowerTen MMA (See~\rsec{microLevel}) in the following Section  --- by the utilization of a platform-specific micro-level lowering algorithm.
This strategy of coupling a target-independent macro-level algorithm for tiling and packing with a target-specific micro-level lowering algorithm can be a roadmap for hardware vendors.
The results in the following section lead to a prediction that, once micro-level lowering is available for \Intel, \AMD, and \PowerNine, the gap between the layered compiler-only solution and high-performance libraries can be bridged.

\subsection{MMA intrinsic}
\label{sec:results-claim2}

The performance of the compiler-only macro kernel can be boosted by exploiting matrix engines.
This study focuses on IBM \PowerTen's MMA and uses the second-silicon version of \PowerTen which has similar hardware characteristics to those that will be featured in commercially available versions but slightly different firmware settings.
Therefore, the reported speedups relative to \BLAS are consistent with the results expected for the commercial version of \PowerTen.
The micro kernels in both \BLAS and \EIGEN were re-engineered by IBM experts to make efficient use of MMA.
The parameters used for these experiments are $\mr=16$, $\nr=8$, and $\kr=128$.

\begin{figure}[t]
  \centering
  \begin{subfigure}{0.49\linewidth}
  \includegraphics[width=0.95\columnwidth]{images/legend-libs-all.eps}
  \includegraphics[width=1.0\columnwidth]{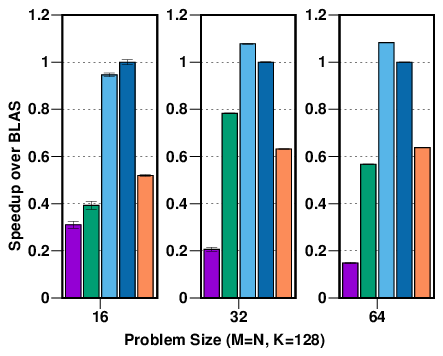}
  \subcaption{Speedup over \BLAS with MMA.}
  \label{fig:speedup-blas-power10-small}
  \end{subfigure}%
  \begin{subfigure}{0.495\linewidth}
  \includegraphics[width=1.0\columnwidth]{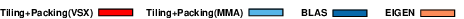}
  \includegraphics[width=1.0\columnwidth]{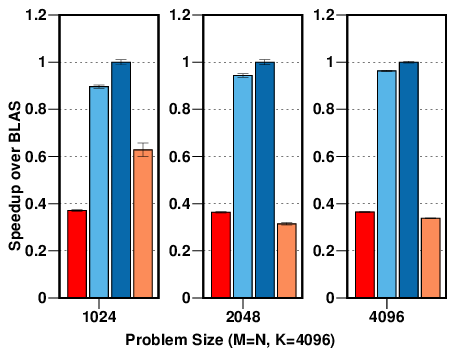}
  \vspace{-0.4cm}
  \caption{VSX vs. MMA performance.}
  \label{fig:vsx-vs-mma}
  \end{subfigure}
  \caption{(a) Speedup over \BLAS of small \sgemm on \PowerTen with MMA; (b) Contrasting VSX and MMA performance of \sgemm kernel on \PowerTen.}
\end{figure}

Different than on other platforms,  \rfig{speedup-blas-power10-small}  shows that for small kernels \TilingPacking results in either better or on-par performance compared to the libraries:
\TilingPacking is over $50\%$ faster than \EIGEN for the problem size $M=N=16$ and over $83\%$ faster for $M=N=32$ and $M=N=64$.
\TilingPacking achieves up to $10\%$ more performance than \BLAS for \sgemms of size $M=N=32$ and $M=N=64$.
With MMA, as \gemm is implement with outer products instead of inner products, columns of matrix A are multiplied by rows of matrix B producing an accumulator result that has partially computed column elements of matrix C -- one per VSR associated with the accumulator.
Therefore, the best access layout for the MMA builtin is both A and C in column-major and B in row-major order.
However, all input matrices are stored, and thus accessed, in column-major order.
Therefore, loaded tiles of matrix B need to be transposed prior to calling the MMA builtin.
For \Intrinsic and \Tiling, this means additional vector shuffle and merge instructions in the micro-kernel code.
The extra instructions in the case of \TilingPacking are not generated as part of the micro level but as part of the macro level algorithm that packs the matrices.
With fewer instructions, the micro-kernel code generated with \TilingPacking runs faster and exhibits better instruction-cache utilization.
In addition, the packing loads act as prefetching loads and reduce access latency for the operands of outer-product instructions.
Moreover, calling a library incurs in performance overheads that are more noticeable on smaller problem sizes.
These results indicate that the proposed macro-level approach coupled with an MMA-specific lowering of \code{llvm.matrix.multiply} can provide better performance than two state-of-the-art libraries.

\rfig{vsx-vs-mma} contrasts the performance of the generic-lowered code, which uses VSX instructions on \PowerTen, with the performance of the new MMA-specific lowered code presented in this paper (\rsec{microLevel}).
The MMA solution is over $2.6\times$ faster than the VSX solution for the largest \sgemm.
The conclusion is that a target-specific lowering can outperform \EIGEN, when both use MMA instructions, for the \sgemm sizes $M=N=2048$ and $M=N=4096$.
In fact, \TilingPacking with VSX matches the performance of \EIGEN with MMA, indicating that the macro-level algorithm made better tiling and packing decisions with compile-time information than \EIGEN.
\TilingPacking provides over $2.6\%$ better performance than \EIGEN for the largest \sgemm.
Furthermore, the MMA-specific lowering algorithm generated code  that closely matches \BLAS performance (up to $96\%$), which itself achieves almost $50$ flops/cycle (almost $80\%$ of peak).
In essence, a compiler-only layered approach built around a compiler intrinsic for matrix multiplication can achieve comparable or better performance than state-of-the-art libraries.
In addition, having a target-specific lowering for the matrix-multiplication instrinsic is key to make best use of vector/matrix units, the higher cache locality, and better utilization of the memory hierarchy delivered by the macro-level algorithm.

\section{Additional Opportunities}
\label{sec:opportunities}

This groundwork implementation of the layered approach in LLVM creates opportunities that will benefit other BLAS kernels and other specialized architectures.\footnote{The implementation will be openly available as an artifact.}

\subsection{Macro-level strategy for other BLAS kernels}
\label{sec:micro-level-other-blas}

\ralg{macrokernel} describes general techniques for blocking, tiling, packing and intrinsic invocation that can be used for other BLAS matrix operations.
As an example, consider \syrtwok, which computes either the lower or upper triangular half of $C \gets \alpha \times A \times B^T + \alpha \times B \times A^T + \beta\times C$.
Matrix $C$ is symmetric while matrices $A$ and $B$ are general.
High performance implementations of \syrtwok partition the matrix $C$ into blocks and use a pair of GEMM operations to update each block.
Both the normal (non-transposed) and transposed versions of matrices $A$ and $B$ are needed.

An equivalent to \ralg{macrokernel} for \syrtwok, will use two calls for packing matrix $B$ (line~\ref{packB}) and two calls for packing matrix $A$ (line~\ref{packA}).
In each case, one of the calls produces a packed version of a block of the matrix and the other call produces a packed version of the transpose of a block
 with different blocks used for the normal and transposed versions.
In the inner loops, the algorithm loads two tiles of $B$ (line~\ref{loadB}) and two tiles of $A$ (line~\ref{loadA}), one tile from the normal block and the other from the transposed, each with different tiles used from the normal and transposed versions.
Finally, the actual computation (line~\ref{multiply}) requires two calls to the \code{llvm.matrix.multiply} intrinsic.
This approach reuses the tiling and packing strategies and allows the use of blocks and tiles of different sizes for matrices $A$ and $B$ to achieve better cache utilization.

The acceleration of other idioms requires changes to our pattern recognition component, KernelFaRer \cite{CarvalhoTACO21}.
For a detailed discussion, please refer to the KernelFaRer paper.
Other idioms can be supported with our macro-level approach.
However, new idioms require simple adjustments to the loop nest generation, mainly to determine a performant loop nesting order.
As other works have shown, tiling and packing are easily automated \cite{Low2016TMS,CaioCGOPacking}.
New compiler intrinsics are required, but adding new intrinsics is a trivial task in compiler development.

\subsection{Targetting other matrix engines}
\label{sec:other-matrix-engines}

\PowerTen MMA is but one of the emerging enhancements conceived to accelerate matrix operations in modern CPUs and GPUs.
For CPUs two other architectural extensions have been announced: Intel's AMX~\cite{IntelISA} and \Arm's ME~\cite{ArmISA}.

Each extension has its own unique characteristics, including the list of supported data types, the set of registers used to operate on matrices, and basic computational operation.
\rtab{featureComparison} shows a brief comparison of the three extensions.
It seems that AMX is more targeted to image-processing and deep learning applications, while both MMA and \Arm's ME can also handle scientific applications.

MMA and Arm's ME are tightly integrated in the processor core while AMX instructions are executed in a more loosely coupled accelerator (TMUL).
An Intel host CPU communicates with an AMX unit via a queue of tiles and accelerator commands~\cite{IntelISA}.
Tile commands load/store tile register data from/to memory addressed by host CPU registers.
All pointer arithmetic and loop control instructions run in the host CPU.
Coherence between accesses from the AMX unit and access from the host CPU is maintained at the main memory level instead of at the cache level~\cite{IntelISA}.

\begin{table}
  \centering
  \caption{Matrix multiplication extensions comparison.}
  \vspace{-0.25cm}
  \footnotesize
  \begin{tabular}{| c | c | c | c |}
    \hline
	  Extension & Types & Registers & Operation\\\hline
      \begin{tabular}{c} \IBM \\ MMA \end{tabular} &
    \begin{tabular}{c}
      \code{i4}, \code{i8}, \code{i16}\\
      \code{bf16}, \code{f16}, \code{f32}, \code{f64}
	\end{tabular} & VSRs + ACCs & \begin{tabular}{c} rank-$k$ update \\ $k = 1,\ldots,8$ \end{tabular}\\ \hline
		\begin{tabular}{c} \Intel \\ AMX \end{tabular} &
    \begin{tabular}{c}
      	\code{i8}, \code{bf16}
    \end{tabular} &
    \begin{tabular}{c}
      Separate register file \\
      $16 \times 64\text{-byte}$ tiles
    \end{tabular} & tile multiplication \\ \hline
	    \begin{tabular}{c} \Arm \\ ME \end{tabular} &
    \begin{tabular}{c}
      	\code{i8}, \code{bf16}, \code{f32}, \code{f64}
    \end{tabular} & Vector registers only & \begin{tabular}{c} $2 \times 2$ \\ matrix operations \end{tabular} \\ \hline
  \end{tabular}
  \label{tab:featureComparison}
  \vspace{-0.15cm}
\end{table}

The approach described in this paper can be used with these other matrix engines.
The macro-level algorithm described in \rsec{macro-level-algorithm} is general and can be lowered to multiple targets supported by LLVM's backend.
The compiler designer must tailor a few lowering operations when moving to a new target.
In particular, the load/store operations for packing (\ralg{macrokernel}, lines~\ref{packB} and \ref{packA}) and for loading and storing tiles (\ralg{macrokernel}, lines~\ref{loadB}, \ref{loadA}, \ref{loadC}, and~\ref{storeC}) must be lowered to architecture-dependent instructions by the LLVM backend.
Lowering to \Arm's ME and \IBM's MMA is straightforward because both matrix engines are tightly coupled accelerators and rely on host CPU vector registers to store input and output data.
On AMX however, the lowering of load and store operations use the\code{TILELOAD*} and \code{TILESTORE*} AMX instructions.
These cause data movements between AMX's tile register and host memory.
There are no instructions to move data between AMX's tile registers and host vector-registers (e.g. AVX-512 registers).
Therefore, for both \gemm and \syrtwok, where partial products are scaled by $\alpha$ and $\beta$ (\ralg{macrokernel}, lines~\ref{scalealpha} and \ref{scalebeta}), tile-register data would have to be stored back to memory and loaded into host vector register for the scaling computations.

\section{Related Work}
\label{sec:relatedWork}

In a seminal work, Goto and Van D. Geijn detail a layered approach to improve cache and vector-register utilization on CPUs ~\cite{goto2008}.
Using this approach, modern linear algebra libraries, such as Eigen and BLAS, achieve high performance on HPC workloads.
Goto and Van D. Geijn show that modelling both L2 cache and TLB --- and not only L1 as considered earlier --- is crucial for cache performance.
That work is seminal because it publicly explained practical strategies for optimal cache and vector register utilization on CPUs; these strategies were previously only available in proprietary libraries.
The layered strategy features two stages:
\begin{inparaenum}
  \item blocking input matrices and packing tiles of these blocks in such a way that tiles lay in main memory in the order that they will be accessed; and
  \item computing a small \gemm at the register level.
\end{inparaenum}
This paper is the first to create a compiler-only code generation for the layered approach and adapts blocking, tiling, and packing to create a data layout that is suitable for computing with MMA and to also improve utilization of the L3 cache.

Gareev et al.~\cite{GareevTACO18} implement tiling and packing within Polly~\cite{polly} without the need for an external library or automatic tuning.
Their approach with a hand-optimized SSE kernel reports performance on par with BLIS on \Intel Sandy Bridge.
When not relying on an assembly kernel, their pass uses the default LLVM vectorizer that delivers only a small speedup over naive code.
The solution proposed in this paper implements both memory optimization and micro kernel in the compiler, not requiring any hand-optimized code.

Uday Bondhugula presents an implementation of the BLAS strategy within the emerging MLIR framework~\cite{uday2020}.
He demonstrates that blocking, packing, register tiling, and unroll+jam yields code that is 34\% slower than OpenBLAS on \Intel's Coffee Lake~\cite{uday2020}.
Bondhugula also implemented a custom vectorization pass, to replace the default LLVM vectorizer, to achieve an additional 40\% performance improvement, thus reaching 91\% of the performance of OpenBLAS.
Our experiments with \Intel, \AMD and \IBM \PowerNine machines also pointed out the weakness of the default LLVM vectorizer.

Carvalho et al. introduce KernelFaRer, a robust pattern recognition system that can identify matrix-multiplication patterns in the LLVM IR level and can replace these with library calls~\cite{CarvalhoTACO21}. While this approach can lead to speedups on the order of 1000s in comparison with non-optimized code, it has the drawback of requiring the integration of libraries into a computer system that may not have it.
Moreover, their experimental results indicate that, for smaller matrices, the overhead of invoking functions in the libraries leads to performance degradations.
The solution in this paper is orthogonal to Carvalho et al.: their pattern recognition can identify GEMM kernels at the intermediate-level representation and then invoke the compiler-only solution presented here.

When presenting the ILLIAC IV, one of the first SIMD machines, Barnes et al. advocated that data parallelism would be crucial for progress~\cite{1968barnes}, citing matrix operations as a critical target~\cite{1968kuck}.
Nearly 50 years later, Barnes' direction culminated in the inclusion of vector extensions in all mainstream CPUs.
Although fast vectorization is powerful, matrix-multiplication performance could be improved further with specialized hardware units.
This possibility is now realized with the introduction of what Domke et al. have dubbed ``matrix engines''~\cite{DomkeIPDPS21}.


Robust performance benchmarking is critical for the evaluation of vector extensions.
While there is extensive performance evaluation of matrix multiplication on vector extensions for  Intel architectures~\cite{2016hassan,2020hemeida,2020alappat}, to the best of our knowledge, similar studies do not exist for the PowerPC or \Arm platforms.
Moreover, the introduction of matrix engines is recent in all platforms and therefore only simulated or theorized performance estimates exist for AMX, SVE, or MMA~\cite{2020poenaru,DomkeIPDPS21}.
Therefore, this work is among the first to present performance evaluation of a matrix engine on actual hardware.

The advent of the ``general purpose'' GPUs quickly saw study and performance analysis of matrix computations~\cite{2001larsen,2004fatahalian}.
This evolved into implementations of matrix multiplications on GPUs: manually~\cite{2011li}, through libraries like BLAS~\cite{2011nath}, and through frameworks such as DistME~\cite{2019han}.
Matrix multiplication is also central to the design of hardware for tensor-operation acceleration such as Google's Tensor Processing Unit~\cite{JouppiISCA17}, Nvidia's Tensor Core~\cite{2018markidis}, and Huawei's Cube Unit~\cite{2019liao}.

Our work proposes a compiler-only approach, thus it does not require profiling data.
Nevertheless, works that automate the process of tuning high-performance implementations of BLAS routines (e.g. ATLAS \cite{ATLAS}) could be coupled with our work.
One possibility is to employ the tuning automation offline to ensure that the compiler-only approach is obtaining the correct tiling and packing factors.
However, as a compiler-only approach, it is more akin to approaches such as code versioning or code generation heuristics, such as the one presented by Rohwedder et al. \cite{CaioCGOPacking}, than offline profiling approaches.

\section{Conclusion}
\label{sec:conclusion}

This work presents a robust solution to the problem of generating efficient code entirely within a compiler targeting multiple architectures, consisting of implementing, in the widely used LLVM compilation framework, the layered approach broadly used in specialized libraries.
A key insight was to create a parameterized algorithm for the tiling and packing layer that only requires the compiler to read the effective sizes of the caches from an existing LLVM pass to determine the appropriate sizes for blocks.
In this approach, a target-specific compilation can use the size of the register file for the target architecture to decide on the most appropriate size for tiles.
Similar to the layered approach used in libraries, the goal of packing is to lay out the tiles in memory in the order in which they will be accessed during the computation to increase locality.
Another essential insight was to use the standard LLVM matrix multiplication intrinsic as the interface between the macro kernel and the micro kernel.
This way, for any target architecture, the specialized code generation at the micro-kernel level only needs to be done once and its performance advantages will benefit any code generation path that uses the same intrinsics.
The performance evaluation, including machines with and without matrix engines, demonstrates the modularity of the design and reveals significant performance gains from the layered approach in multiple architectures.
The experimental evaluation indicates that the macro-level algorithm, coupled with a generic intrinsic lowering, achieves more than $22\times$ better performance than PluTo, a widely used compiler-only polyhedral optimizer.
The new compiler-only approach generates code that matches \EIGEN performance and is only $34\%$ slower than \BLAS on \PowerNine.
An MMA-specific implementation results in more than $2.6\times$ the performance of the VSX micro kernel, is over $83\%$ faster than \EIGEN, and achieves up to $96\%$ of \BLAS peak performance for large \sgemms, even when these libraries are engineered to also benefit from MMA.

%

\bibliography{refs.bib}

%
%

\end{document}